\documentclass{aa}
\usepackage{graphicx} 
\usepackage[varg]{txfonts}
\usepackage{natbib,twoopt}
\usepackage{adjustbox}
\usepackage{soul}
\usepackage{float}
\usepackage{placeins}
\usepackage{longtable} 
\usepackage{lscape}    
\usepackage[breaklinks=true]{hyperref} 
\bibpunct{(}{)}{;}{a}{}{,}             
\makeatletter
  \newcommandtwoopt{\citeads}[3][][]{\href{http://adsabs.harvard.edu/abs/#3}%
    {\def\hyper@linkstart##1##2{}%
     \let\hyper@linkend\@empty\citealp[#1][#2]{#3}}}
  \newcommandtwoopt{\citepads}[3][][]{\href{http://adsabs.harvard.edu/abs/#3}%
    {\def\hyper@linkstart##1##2{}%
     \let\hyper@linkend\@empty\citep[#1][#2]{#3}}}
  \newcommandtwoopt{\citetads}[3][][]{\href{http://adsabs.harvard.edu/abs/#3}%
    {\def\hyper@linkstart##1##2{}%
     \let\hyper@linkend\@empty\citet[#1][#2]{#3}}}
  \newcommandtwoopt{\citeyearads}[3][][]%
    {\href{http://adsabs.harvard.edu/abs/#3}
    {\def\hyper@linkstart##1##2{}%
     \let\hyper@linkend\@empty\citeyear[#1][#2]{#3}}}
\makeatother

\title{Constraining young massive cluster properties with radio-continuum observations: The Arches cluster}

\author{
        M. Cano-Gonz\'alez
         \inst{1}
          \and
        R. Sch\"odel
         \inst{1}
          \and
          A. Alberdi
         \inst{1}
         \and
          F. Najarro
         \inst{3}
         \and
         J. Moldón
         \inst{1}
          \and
          M. P\'erez-Torres
         \inst{1,2}
         \and
         F. Nogueras-Lara
         \inst{1}
}
\institute{
           Instituto de Astrof\'isica de Andaluc\'ia (CSIC),
           Glorieta de la Astronom\'ia s/n, 18008 Granada, Spain.
           \email{mcano@iaa.es}
           \and
           School of Sciences, European University Cyprus, Diogenes street, Engomi, 1516 Nicosia, Cyprus
           \and
           Centro de Astrobiolog\'ia, CSIC-INTA, Ctra de Torrej\'on a Ajalvir km 4, 28850 Torrej\'on de Ardoz, Madrid, Spain
}
\date{}
\begin{document}

  \abstract
   {The Arches cluster, located in the Galactic Centre (GC) is one of the best astrophysical laboratories to study the properties of massive stars and young massive clusters (YMCs). However, despite having been observed for decades at different wavelengths, several fundamental parameters of the Arches cluster remain uncertain.}
   {Our goal is to constrain key cluster parameters (cluster age, mass, and initial mass function, IMF) by comparing the observed stellar radio flux density distribution of the Arches cluster to those derived from a set of synthetic clusters.}
   {We use the deep X-band (10 GHz) Very Large Array data from our previous radio continuum study of the Arches cluster. We model each simulated cluster with three parameters: age, mass, and IMF slope. We use three different stellar evolutionary models: GENEC, PARSEC, and MIST at two different metallicities, solar ($Z=0.014$) and super-solar ($Z=0.020$). We run Markov-chain Monte-Carlo simulations for each model/metallicity combination in order to explore parameter space.}
   {All models and metallicities return preferred ages in the $2\lesssim t_{\rm age}/{\rm Myr}\lesssim 3$ range. Using radio data alone we obtain an IMF slope of $\alpha_{\rm IMF}=-1.85^{+0.28}_{-0.20}$, averaged over all models, where uncertainties are dominated by the degeneracy between cluster mass and IMF slope. If we use the IMF slope from previous infrared studies as prior, the cluster mass distributions peak at $\sim2.7\times10^4\, M_\odot$ and we can establish a lower limit at $\gtrsim2\times10^4 M_\odot$ for the Arches cluster mass.}
   {Radio continuum observations of their  most massive stars can be used to constrain YMC parameters. In the case of the Arches cluster, age can be determined regardless of prior spectroscopic information, which can be useful to characterise  newly discovered YMCs in the GC. Furthermore, our results support the idea that a top-heavy IMF may be preferred in the GC or in YMCs in general.}

   \keywords{Stars: massive --
                Galaxy: centre --
                Open clusters and associations --
                Radio continuum: stars
               }
   \maketitle
%

\section{Introduction}

Massive stars $(M_{\rm ini}\gtrsim 8\, M_\odot)$ shape the interstellar medium with their intense UV radiation and stellar winds, and play a key role in chemical enrichment and energy injection into the interstellar medium because of their demise as supernovae \citep[e.g.~][]{Woosley2002_review}. Massive stars are rare targets, live shortly ($\lesssim10$ Myr), and are typically located at kiloparsec distances, where extinction is more likely to be stronger. Contrary to their low-to-intermediate mass counterparts, the evolution of massive stars is still relatively poorly understood, especially their post-main sequence stages \citep[e.g.~][]{Langer2012_review, Martins_Palacios2013}. This is in part due to the fact that most massive stars lie in binary or higher order multiple systems, in which interactions between members can drastically affect their evolution via phenomena like mass transfer, common envelopes, and rejuvenation \citep[e.g.~][]{Marchant2024_review}. Furthermore, mass-loss rates drastically increase during the Wolf-Rayet (WR) stage, and eruptive episodes of mass-loss can occur during the luminous blue variable (LBV) phase of the most massive stars ($M_{\rm ini}\gtrsim 60\, M_\odot$), as observed in systems like $\eta$-Carinae \citep{Smith2008}. In addition, the exact evolutionary pathway (and lifetime of each WR phase, \citealp[e.g.~][]{Georgy2012}) of a massive star is highly sensitive to its initial mass, with the most massive stars ($M_{\rm ini}\gtrsim 60\, M_\odot$) skipping the red supergiant phase entirely, and going through the following stages: OIa\textsuperscript{(+)}/WNh (blue supergiant/hydrogen-rich WR) $\rightarrow$ LBV $\rightarrow$ WN (nitrogen-rich WR)$\rightarrow$ WC (carbon-rich WR) $\rightarrow$ WO (oxygen-rich WR) $\rightarrow$ remnant; (\citealp[e.g.~][]{Groh2014}, although such sequence can change depending on metallicity \citealp[e.g.~][]{Szecsi2022}, and/or binary interactions \citealp[e.g.~][]{Marchant2024_review}).

Young massive clusters (YMCs) are the best laboratories to study the properties and evolution of massive stars, as they usually harbour dozens of them formed quasi-coevally. So far, only a few YMCs have been discovered in the Milky Way \citep[e.g.~][]{Figer2004,Clark2005,Clark2005_Wd1,Ascenso_2007}, and most of those observed outside the Galaxy (with some exceptions belonging to the Magellanic clouds \citealp[e.g.~][]{Crowther2016}), are too far for current instruments to resolve individual cluster members, and rely on integrated light measurements to study their stellar populations.

The Arches cluster, located $\sim8$ kpc away in the Galactic centre (GC), is an excellent example of a YMC. Despite having been studied for decades at different wavelengths \citep[e.g.~][]{Lang2005,Wang2006,Clark_I}, the Arches cluster still raises questions regarding its age \citep[e.g.~][]{Schneider2014, Clark_IV}, mass \citep[e.g.~][]{Harfst2010,Clarkson2012, Habibi2013}, and its initial mass function (IMF) \citep[e.g.~][]{Kim2006,Harfst2010,Hosek2019}. Indeed, the massive end of the IMF in YMCs plays a crucial role in stellar remnant populations, and Galactic and chemical evolution in general \citep{Bastian2010_review}.

One of the main advantages of observing in radio frequencies is that interstellar extinction is negligible, avoiding the systematic uncertainties related to the assumption of an extinction law (in the infrared, wavelength at which most stellar population studies in the GC are carried out \citealp[e.g.~][]{GALACTICNUCLEUS_III,Hosek2022}). Assumptions on stellar reddening affects fundamental stellar parameters such as luminosities and mass-loss rates \citep[e.g.~][]{Hosek2019,Clark_IV}. 

In this work we explore the potential of radio continuum observations to derive YMC parameters. In particular, we use the X-band (central frequency of 10 GHz) continuum data from \citet{Cano-Gonzalez2024} to infer key properties of the Arches cluster via Bayesian modelling. Our simulations mostly return ages in the $2-3$ Myr range, with IMF slopes consistent with a top-heavy IMF in the GC environment. Furthermore, when combining our results with recent infrared studies, we can establish a lower limit to Arches cluster mass at $\gtrsim20\,000\, M_\odot$.

\section{Sample and data}\label{Sect_Obs}
In this work, we used the \textit{Karl G. Jansky Very Large Array} (VLA) data from \citet{Cano-Gonzalez2024}. In particular, we used the flux densities extracted from their deep X-band image, as it provides the most complete radio-stellar catalogue of the cluster to date (with a total of 23 radio-stars detected, including all the WNh population of the Arches). We refer to this article for details regarding data reduction, image deconvolution, flux density extraction, spectral index determination, and mass-loss estimation. Table \ref{Table_walpha} shows the sample of radio-stars used, as well as their spectral type and average spectral index $(\alpha,\, \rm where\, S_\nu \propto \nu^\alpha)$.

\setlength{\tabcolsep}{5pt}
\begin{table}
\caption{Radio-stellar sample of the Arches cluster.}
\label{Table_walpha}
\centering
\begin{tabular}{l c c c}
\hline \hline
ID & type & $\bar{\alpha}_{4-12\rm \, GHz}$ &  $S_X$ (mJy) \\
\hline
F6 & WN8-9h + ? & $-0.11 \pm 0.01$ & $2.277\pm0.069$\\ 
F7 & WN8-9h & $0.71 \pm 0.02$ & $0.422\pm0.015$\\ 
F3 & WN8-9h & $0.51 \pm 0.03$ & $0.264\pm0.009$\\ 
Dong19 & WN8-9h & $0.96 \pm 0.04$ & $0.239\pm0.008$\\ 
F4 & WN7-8h & $0.86 \pm 0.08$ & $0.237 \pm 0.009$\\ 
F5 & WN8-9h & $0.62 \pm 0.04$ & $0.227 \pm 0.008$\\ 
F8 & WN8-9h & $0.61 \pm 0.05$ & $0.216 \pm 0.014$\\ 
Dong96 & WN8 & $0.96 \pm 0.08$ & $0.187 \pm 0.009$\\ 
F1 & WN8-9h & $0.43 \pm 0.10$ & $0.180 \pm 0.008$\\ 
F2 & WN8-9h + O5-6Ia$^+$ & $0.58 \pm 0.12$ & $0.146 \pm 0.006$\\ 
F9 & WN8-9h & $0.81 \pm 0.14$ & $0.139 \pm 0.006$\\ 
F12 & WN7-8h & $0.05 \pm 0.30$ & $0.086 \pm 0.006$\\ 
B1 & WN8-9h & $0.82 \pm 0.36$ & $0.069 \pm 0.005$\\ 
F14 & WN8-9h & $-0.75 \pm 0.37$ & $0.058 \pm 0.009$\\ 
F16 & WN8-9h & -- & $0.030 \pm 0.006$ \\
\hline
F19 & O4-5Ia & $-0.38 \pm 0.05$ & $0.291\pm0.010$ \\ 
F18 & O4-5Ia$^+$ & $-0.18 \pm 0.09$ & $0.157 \pm 0.007$\\ 
F26 & O4-5Ia & $-0.47 \pm 0.08$ & $0.152 \pm 0.008$\\ 
F55 & O5.5-6III & $-0.99 \pm 0.60$ & $0.038 \pm 0.011$\\ 
F10 & O7-8Ia$^+$ & -- & $0.029 \pm 0.004$ \\
F17 & O5-6Ia$^+$ & -- & $0.021 \pm 0.005$ \\
F15 & O6-7Ia$^+$ + ? & -- & $0.017 \pm 0.004$ \\
F28 & O4-5Ia & -- & $0.013 \pm 0.003$ \\
\hline \hline
\end{tabular}
\tablefoot{Radio-stars sorted by stellar type and flux density. Spectral types from \citet{Clark_I, Clark_IV} or \citet{Dong2011}. Mean spectral indices obtained from the C- and X-band observations of \citet{Cano-Gonzalez2024}. Flux densities extracted from the deep X-band image of \citet{Cano-Gonzalez2024}.}
\end{table}

Our radio-stellar catalogue of the Arches cluster shows a clear correlation between spectral type and radio emission type. Namely, most WNh stars show spectral indices consistent with thermal radio emission, suggesting that the line-driven winds may be the dominant mechanism behind their observed flux densities. On the other hand, the negative measured spectral indices of O-type supergiants indicate that non-thermal (synchrotron) emission dominates their observed radio emission with the VLA. 

The brightest radio-star, F6, is the only clear exception to the correlation between WNh stars and thermal radio emission. Its spectral indices from the observations in \citet{Cano-Gonzalez2024} indicate a clear flat-to-inverted spectrum, inconsistent with the canonical thermal value of $\alpha\sim0.6$. Additionally, F6 is a variable radio source, and the observations in \citet{Cano-Gonzalez2024} do not cover frequency ranges in which non-thermal emission can be safely ignored ($\nu\gtrsim50$ GHz). Therefore, we could not reliably estimate the thermal component of its total emission. For this reason, we decided to exclude F6 from our observed sample.

In addition, the spectral indices of F12 and F14 are compatible with non-thermal emission in some epochs \citep[see][Table 1]{Cano-Gonzalez2024}, but their relatively faint flux densities result in high spectral index uncertainties. Furthermore, F14 position coincides with some non-thermal extended emission near cluster centre, potentially biasing the flux densities extracted in the C-band observations, resulting in more negative spectral index values.

\section{Methodology and analysis}\label{Sect_methods}
\subsection{Initial assumptions: thermal vs non-thermal emission}\label{subsect_initassump}
Motivated by the $\alpha$-spectral type bi-modality discussed in the previous section, in the simulations described in the following, we assumed that WR-type stars (including Arches-like WNh stars) are (the) only thermal emitters; and that O-type, non-WR stars emit only non-thermally. In fact, if we use the \citet{Wright_Barlow1975} formula to derive thermal fluxes for the O-type giants from the \citet{DeBecker2013} colliding-wind binary catalogue, we can see that the non-thermal emission dominates the observed fluxes by a factor of $10^{4-5}$ in OB-OB systems. Therefore, we can safely ignore the thermal contributions from O supergiants (and even more so main sequence stars).

Consequently, when extracting the flux density distribution (FDD) of a synthetic cluster, we considered the two different types of emission separately: $S_\nu^\mathrm{cluster}=S_\nu^\mathrm{WR}(\mathrm{thermal)} + S_\nu^\mathrm{OB}(\mathrm{non\text{-}thermal})$.

\subsection{Simulating radio continuum emission from synthetic clusters}
We used the Stellar Population Interface for Stellar Evolution and Atmospheres (SPISEA) python package \citep{Hosek2020} to create synthetic clusters. We assumed that all of them were located at the GC distance of 8\,kpc \footnote{We refrain from investigating the impact of this systematic uncertainty as it affects all Arches members equally and the distance from the central black hole to the Arches cluster is not precisely defined.}.

Along with the massive stellar evolutionary models included in the stand-alone SPISEA version, that assume solar metallicity for all models but MIST, we added one extra metallicity ($Z=0.020$) for the GENEC \citep{Ekstrom2012,Georgy2012,Yusof2022}, MIST \citep{Choi2016}, and PARSEC models \citep{Bressan2012, Chen2015} as well as expanded the age range covered by the PARSEC isochrones, so that they would cover the youngest ages that correspond to the Arches cluster. We used these three evolutionary models because they were the only ones included in SPISEA that covered the mass range of radio-stars $(M_*\gtrsim40\, M_\odot)$. Furthermore, in order to properly characterise and classify WR stars, we modified the \texttt{synthetic.py} and \texttt{evolution.py} scripts included in the SPISEA distribution so that we could extract mass-loss rates and surface abundances from the isochrones\footnote{The author is willing to share the modified scripts upon reasonable request.}. Namely, we extracted hydrogen, helium, carbon, nitrogen, and oxygen surface abundances ($X_{\rm H}$, $X_{\rm He}$, $X_{\rm C}$, $X_{\rm N}$, and $X_{\rm O}$, respectively).

\subsubsection{Thermal emission}\label{subsect_thermalemission}
Different evolutionary models define the WR stage differently. While the classification always depends on hydrogen surface abundance $X_{\rm H}$, the exact threshold varies among models. For example, GENEC (MIST) models impose $X_{\rm H}<0.3$ ($X_{\rm H}<0.4$) for a star to be considered WR. However, many of the Arches WR stars are still in their hydrogen-burning phase (WNh), and may present higher H-abundances, up to $X_{\rm H}\gtrsim0.5$. In fact, if we were to take the GENEC or MIST criteria, only a handful of the Arches WNh stars would be considered WR by the code, despite showing clear spectroscopic features of WR stars \citep[e.g.~][]{Martins2008,Clark_I}. 

Therefore, in order to include hydrogen-rich WR stars in our simulated sample, we impose the following criteria: $0.2<X_{\rm H}<0.6$ and $M_{\mathrm{ini}}>80\, M_\odot$ where $M_{\mathrm{ini}}$ is the initial stellar mass. The upper limit of $X_{\rm H}<0.6$ is imposed so that hardly evolved O-type stars are not classified mistakenly as WR. The lower limit of $X_{\rm H}>0.2$ is set so that the helium number values obtained from the synthetic stars resemble those of the Arches cluster \citep{Martins2008}. The rest of the WR subtypes (WNL, WNE, WC, and WO) are classified following \citet{Georgy2012} criteria, which take into account surface abundances and effective temperatures. In short, from the surface abundance standpoint, a star is classified as WNL if it still shows traces of hydrogen in its surface ($X_{\rm H}>10^{-5}$), and it is classified as WC if it is hydrogen-depleted ($X_{\rm H}<10^{-5}$), and $X_{\rm C}>X_{\rm N}$.

\begin{figure}
  \centering
  \includegraphics[width=\hsize]{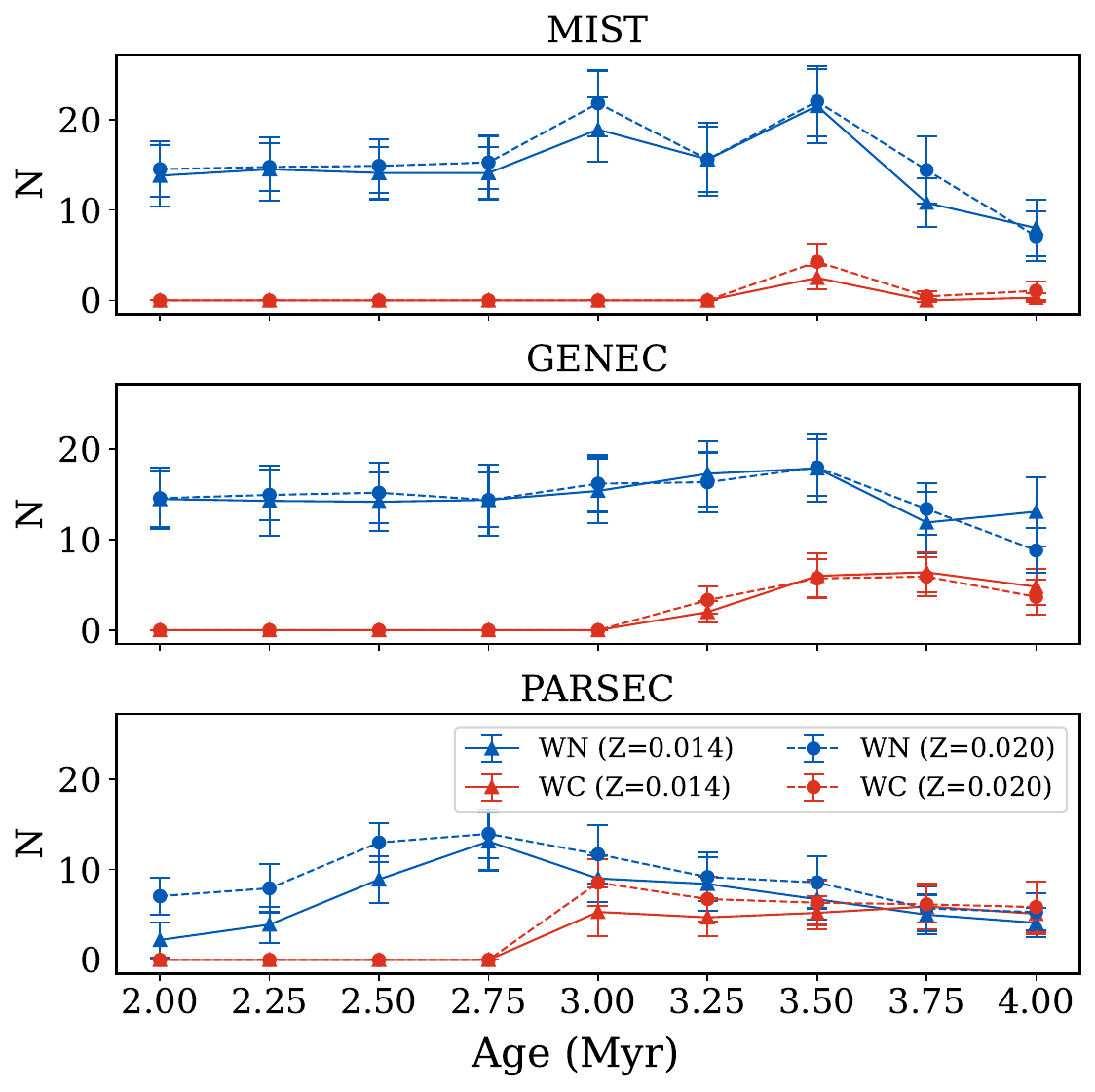}
     \caption{Average WN and WC counts for a $20\,000\, M_\odot$ cluster with $\alpha_{\mathrm{IMF}}=-1.8$ \citep{Hosek2019} and two different metallicities: solar $(Z=0.014)$ and super-solar $(Z=0.020)$. Each panel shows the results obtained for a different evolutionary model. We computed mean counts  per cluster, age, and metallicity from 50 realisations. As reference, the Arches cluster hosts 15 known WN(h) stars and no WC star has been observed.}
        \label{Fig_WRcounts}
\end{figure}

With these criteria there is an overlap between the WNh and WNL subtypes (for example, a star with $0.2<X_{\rm H}<0.3$ and $M_{\rm ini}>80\, M_\odot$ would meet both WNh and WNL criteria). Therefore, for our synthetic sample of thermally emitting radio-stars, we considered both the stars meeting the aforementioned WNh criteria and those characterised as WNL by the evolutionary model. This way we include all hydrogen rich WR stars in the our simulated sample. In the following, for simplicity's sake, we will refer to both WNh and WNL as WN, as they are only different in nomenclature.
Figure \ref{Fig_WRcounts} shows that our criteria lead to consistent numbers of WN stars for all three evolutionary models used.

With the synthetic WN stars  identified, we determined their thermal radio emission.  To do so, we used the \citet{Wright_Barlow1975} prescription, which relates stellar mass-loss with free-free (thermal) emission. As in \citet{Cano-Gonzalez2024}, we used the following  parameters, assuming a fully ionised stellar wind: mean ionic charge and mean number of electrons per ion  $Z_{\rm wind}=\gamma=1$;  mean molecular weight, $\mu$ from the Helium number approximation \citep[see, e.g.][Eq. 5]{G-C2021}; terminal wind velocities ($v_\infty$) are computed from the average of all Arches WNh stars \citep[][Table 2]{Martins2008}, and an electron temperature, $T_e= 10^{4}$ K.

Thermal radio emission depends significantly on the particular mass-loss prescription applied, and how the prescription treats clumping. While all models use the mass-loss prescription of \citet{Nugis&Lamers2000} for the WR stage, the GENEC models also include the prescription by \citet{Grafener&Hamann2008} during the hydrogen-rich WR phase. Upon further investigation, we found an important difference in how each prescription defines clumping in the outer layers of the stellar wind. For instance, \citet{Grafener&Hamann2008} define the outer wind as the distance at which $\tau_R<0.35$, where $\tau_R$ is the Rosseland optical depth. This, in the case of a massive WN star, can be around 1.5 to 3 stellar radii away, and therefore their assumption of a constant maximum clumping factor\footnote{Where $f_{\rm cl}$, the clumping factor, is defined as $f_\mathrm{cl}=\langle \rho^2 \rangle/\langle \rho\rangle^2 \geq1$. The mass-loss rates are therefore scaled by clumping in the following manner: $\dot{M}_{\rm smooth}=\dot{M}_{\rm clumped} \sqrt{f_{\rm cl}}$.} of $f_{\rm cl}^{\rm GH08}=10$ is justified, as some studies show that clumping reaches a maximum value up to $10\,R_*$ and decreases continuously towards larger atmospheric heights \citep[e.g.~][]{Runacres&Owoki2002}. On the other hand, \citet{Nugis&Lamers2000} used sub-mm to FIR data from \citet{Nugis1998} to derive clumping-corrected mass-loss rates. At these frequencies, the thermal emission is formed further away from the stellar surface, $R_X\gtrsim100\,R_*$, e.g. \citealt{Wright_Barlow1975}, and clumping is less prominent.

Using the WNL sample from \citet[][see their Table 6]{Nugis1998}, we can estimate an average clumping factor of $f_{\rm cl}^{\rm N98}\approx1.75$, considerably lower than the maximum value adopted by \citet{Grafener&Hamann2008}. Since the results from \citet{Nugis1998} are derived from a study that covers the wavelength range of our data, we decided to scale the mass-loss values from \citet{Grafener&Hamann2008} by $\sqrt{1.75/10}$, in the range were they are applicable, that is, for massive stars in the $0.3<X_{\rm H}<0.6$ range. Although this range is not explicitly shown in the main GENEC models \citep{Ekstrom2012,Georgy2012,Yusof2022}, we found a difference in mass-loss behaviour at around $X_{\rm H}\sim0.3$, which we attributed to the change in mass-loss prescription from \citet{Grafener&Hamann2008} to \citet{Nugis&Lamers2000}. Therefore, in the simulations performed with the GENEC models, we used the scaled mass-loss values for WN with $X_{\rm H}>0.3$, and the standard (returned by the model) values for WN stars with $X_{\rm H}<0.3$. Furthermore, without this scaling, GENEC mass-loss rates (and subsequently thermal flux densities), were underestimated by a factor of 2-4 when compared to the observed upper limits derived for the Arches WNh stars, that span the $(1-6)\times10^{-5}\, M_\odot\, \rm yr^{-1}$ range, approximately \citep{Cano-Gonzalez2024}. We note that the MIST and PARSEC models did not require additional clumping scaling, as the mass-loss rates provided by these models agreed well with the observed range of the Arches WNh stars, as shown by Table \ref{Table_meanMdot}. The values reported in Table \ref{Table_meanMdot} also agree well with the Galactic WN sample of \citet{Hamann2019}.

\def\arraystretch{1.2}
\setlength{\tabcolsep}{5pt}
\begin{table}
\caption{Average mass-loss rates for the three evolutionary models and the thermal radio-stars of the Arches cluster.}
\label{Table_meanMdot}
\centering
\begin{tabular}{l c}
\hline \hline
Model/sample & $\dot{M}\, (\times10^{-5}M_\odot\,\rm yr^{-1})$  \\
\hline
GENEC & $2.54 \pm 0.59$ \\
MIST & $5.3 \pm 2.4$ \\ 
PARSEC & $4.2 \pm 2.9$ \\ 
\hline
Arches & $3.15\pm1.55$ \\ 
\hline \hline
\end{tabular}
\tablefoot{The mean mass-loss value for the Arches thermal radio-stars was obtained by averaging the values in the columns corresponding to the X-band observations in \citet[][Table 3]{Cano-Gonzalez2024}. The mean values for each evolutionary model were obtained by averaging over fifty, 2.5 Myr clusters per model, at solar metallicity. The uncertainties denote standard deviations.}
\end{table}

It is established that very massive stars (VMS, $M_{\rm ini}\gtrsim100\,M_\odot$) can suffer from enhanced mass-loss due to proximity to the Eddington scattering limit \citep[e.g.~][]{Vink2015_VMS}. Fortunately, all models used in this work include mass-loss enhancement for VMS. In particular, PARSEC models use the enhanced prescriptions from \citet{Vink2011}, and both MIST and GENEC incorporate mass-loss enhancement as a by-product of stellar rotation in VMS \citep{Choi2016,Yusof2022}.

\subsubsection{Non-thermal emission}

Non-thermal processes in colliding-wind binaries can produce bright emission at the frequencies sampled by the VLA C- and X-band (central frequencies of 6 and 10 GHz, respectively). However, there is no clear connection between stellar parameters and non-thermal flux, as the synchrotron emission depends not only on the strength of the winds, but also on orbital parameters \citep[e.g.][]{Dougherty2005, DeBecker2007}. The latter are sampled by SPISEA following \citet{DK2013_review}.

\begin{figure}
  \centering
  \includegraphics[width=\hsize]{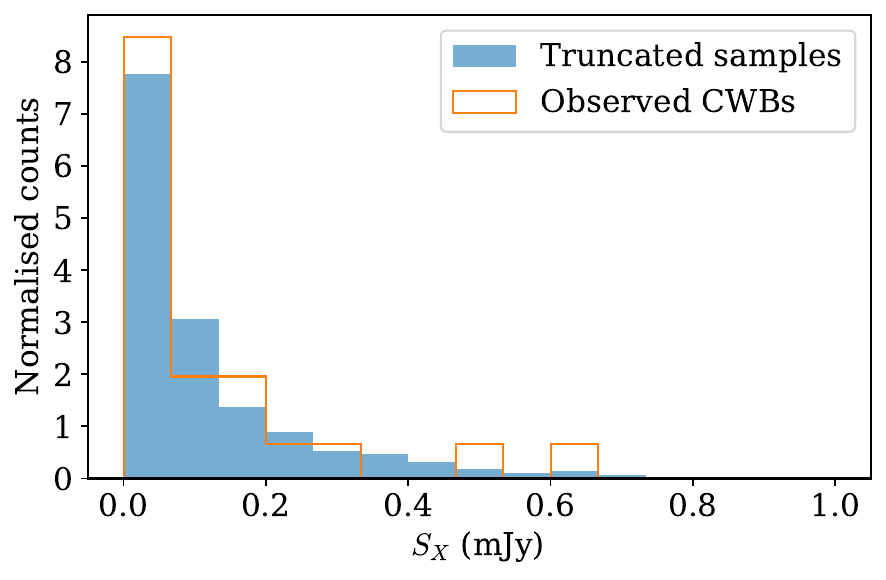}
     \caption{\textit{Blue:} Counts drawn from a truncated $\Gamma$ distribution ($1\,500$ iterations with lower limit of 0.012 mJy, the 5$\sigma$ rms of the observations used here; \citealp{Cano-Gonzalez2024}). \textit{Orange:}  Observed flux densities from the 23 OB supergiants from \citet{DeBecker2013} and \citet{Cano-Gonzalez2024,Cano-Gonzalez2025}.}
    \label{Fig_CWBdist}
\end{figure}

We combined the OB-OB binary supergiant catalogue from \citet{DeBecker2013} with the OB radio-stars identified in the Arches and Quintuplet clusters \citep{Cano-Gonzalez2024,Cano-Gonzalez2025}, that we also assume to be colliding wind binaries. With the  spectral indices and distances given in  \citet{DeBecker2013}, we transformed the flux densities of their sample to the corresponding central frequency of 10 GHz of the VLA X-band at the assumed distance of 8 kpc to the GC. The orange histogram in Fig.\,\ref{Fig_CWBdist} shows the flux distribution of the 23 OB-OB colliding wind binaries in our combined sample.

We then modelled the distribution with  a Gamma function of the form:
\begin{equation}
    f(S_X; \alpha_\Gamma, \theta_\Gamma) = \frac{1}{\Gamma(\alpha_\Gamma)\, \theta_{\Gamma}^{\alpha_\Gamma}} \, S_X^{\alpha_\Gamma - 1} \, e^{-S_X/\theta_\Gamma}, \quad S_X > 0.
\end{equation}
The best fit parameters to the observed distribution are $\alpha_\Gamma=0.19^{+0.21}_{-0.18}$ (shape parameter) and $\theta=0.29\pm0.09$ mJy (scale parameter), where the uncertainties were obtained via bootstrapping. The rather large uncertainties of the $\alpha_\Gamma$ parameter are most likely due to the scarce sample of only 23 observed OB-OB systems. The blue histogram in Fig.\,\ref{Fig_CWBdist} shows the samples drawn from the best-fit $\Gamma$ distribution, truncated at the $5\,\sigma$ detection limit of the deep X-band observations of \citet{Cano-Gonzalez2024}.

\begin{figure}
  \centering
  \includegraphics[width=\hsize]{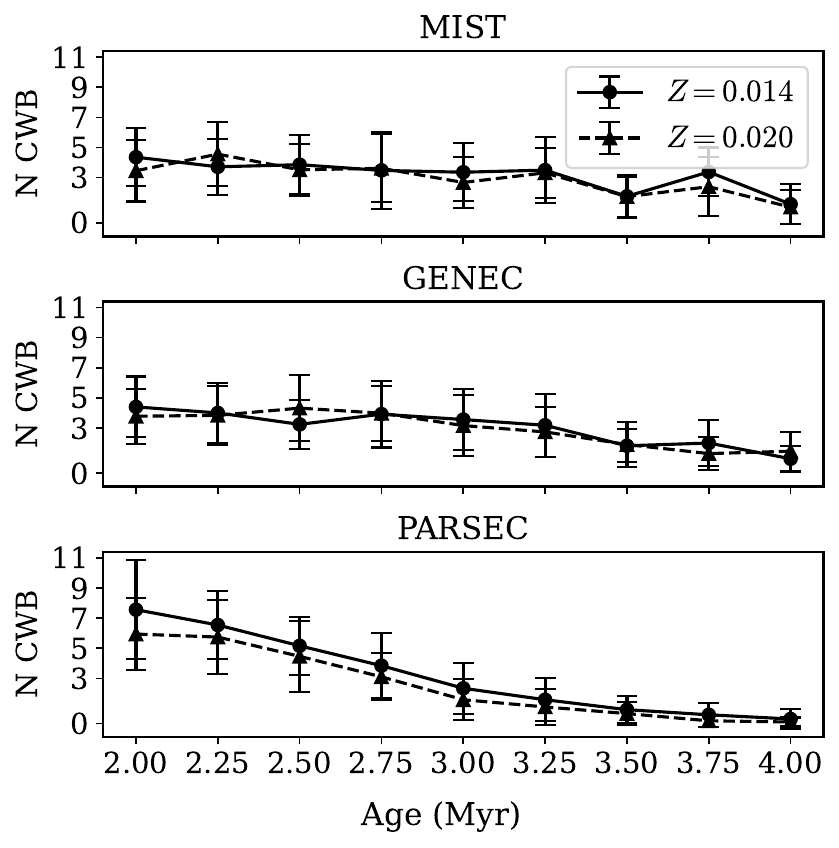}
     \caption{Average counts of expected CWBs per 50 simulated clusters for the three evolutionary models considered. All clusters host a total of $20\,000\, M_\odot$ in stellar mass, and were created with a top-heavy IMF slope of $\alpha_\mathrm{IMF}=-1.8$. As reference, in the Arches cluster, we detected a total of 7 OIa supergiants with the VLA.}
    \label{Fig_CWBcounts}
\end{figure}

In order for a synthetic OB-OB supergiant system to be considered as part of the non-thermal component of the cluster FDD, it must meet the following criteria: both the primary and companion initial masses ($M_{\rm ini}$) must be larger than $40\, M_\odot$, their effective temperatures ($T_{\rm eff}$) larger than $25\,000$ K, and they must not be WR stars.
In addition, we took into consideration the fact that not all O supergiants in the Arches (or Quintuplet) are detected in our radio observations. In particular, only $7/32\sim22\%$ of the Arches O supergiants (types O Ia$^{(+)}$) are clear radio detections. Therefore, even if a synthetic multiple O-type system met the previous criteria, only $22\%$ of them were considered to have properties that make them observable. This fraction of systems was then used to draw a flux density value that is added to the simulated cluster's FDD. This criterion can be thought of as selecting only those OB binary systems whose orbital separation is large enough to allow particle acceleration (producing synchrotron emission). A lower limit of two weeks to the orbital period is suggested in \citet{DeBecker2007,DeBecker2013}, however, when examining the orbital parameters of our synthetic stars meeting the aforementioned $M_{\rm ini}$ and $T_{\rm eff}$ criteria, a minimum orbital period of $\gtrsim 100$ days better reproduced the observed O supergiant radio-stars in the Arches cluster. Figure \ref{Fig_CWBcounts} shows the mean number of synthetic non-thermal OB systems for different cluster ages at two metallicities.

\subsection{MCMC simulations}
\begin{figure}
  \centering
  \includegraphics[width=\hsize]{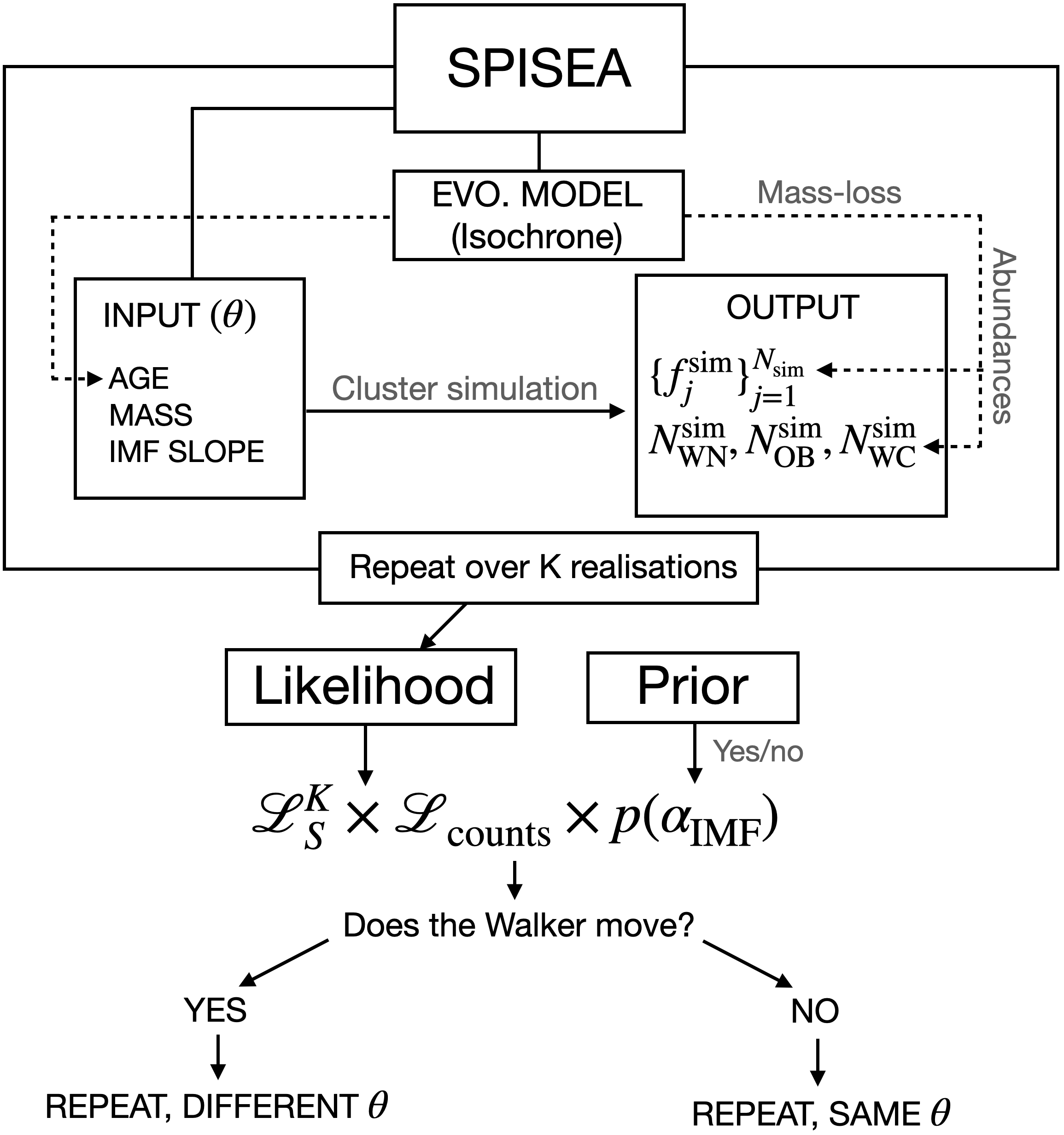}
     \caption{Flow chart of a single iteration of the MCMC simulation. Each walker undergoes the same process in parallel. A realisation is essentially defined as a cluster simulation. For more details on these concepts, see Sect. \ref{subsubsect_MCMCsetup}.}
    \label{Fig_flowchart}
\end{figure}

We explored the preferred regions of the parameter space with Markov-chain Monte-Carlo (MCMC) simulations. Figure \ref{Fig_flowchart} shows an schematic flow chart of the process. In the following subsections, we explain in detail how we modelled the clusters and built our simulations.

\subsubsection{Forward model}

We modelled each cluster using the parameter vector
\begin{equation}
\boldsymbol{\theta}
=
\left(
t_{\rm age},\;
M_{\rm cl},\;
\alpha_{\rm IMF}
\right),
\end{equation}
where $t_{\rm age}$ is the cluster age (in years), $M_{\rm cl}$ is the cluster mass (in $M_\odot$), and
$\alpha_{\rm IMF}$ the IMF slope,
defined so that $dN/dM_* \propto M_{*}^{\alpha_{\rm IMF}}$, where $M_*$ is the initial mass of a given star.

For a given $\boldsymbol{\theta}$, each realisation (cluster simulation) returns: 1) The FDD of the synthetic cluster: an array of X-band (10 GHz) simulated flux densities (in mJy) that is
subject to the same detection threshold as the deep X-band observations ($S_X>0.012$ mJy); 2)
the number of simulated WN stars ($N_{\rm WN}$); 3) the number of synthetic OB colliding-wind
binary systems ($N_{\rm OB}$); and 4) the number of synthetic WC stars ($N_{\rm WC}$).

\subsubsection{Likelihood}

Given the set observed X-band flux densities\footnote{As previously mentioned, we decided to exclude F6 from the observed sample. Forcing F6 into the simulations had a negative impact on acceptance rate and convergence, presumably because the simulated clusters find it difficult to replicate such a bright radio-star ($\gtrsim2.3$ mJy) within the thermal/non-thermal dichotomy.} from \citet{Cano-Gonzalez2024} $\{f_i^{\rm obs}\}_{i=1}^{N_{\rm obs}}$, and a simulated cluster containing $N_{\rm sim}$ fluxes $\{f_j^{\rm sim}\}_{j=1}^{N_{\rm sim}}$, we define
the likelihood for a single observed flux density as
\begin{equation}
p(f_i^{\rm obs} \mid \boldsymbol{\theta})
=
\frac{1}{N_{\rm sim}}
\sum_{j=1}^{N_{\rm sim}}
\mathcal{N}
\left(
f_i^{\rm obs} \mid f_j^{\rm sim}, \sigma_{i}
\right),
\end{equation}
where $\mathcal{N}$ denotes a Gaussian probability density 
\begin{equation}
    \mathcal{N}
\left(
f_i^{\rm obs} \mid f_j^{\rm sim}, \sigma_{i}
\right)
= \frac{1}{\sqrt{2\pi}\,\sigma_i}\exp \left[ -\frac{(f_i^{\rm obs} - f_j^{\rm sim})^2}{2\sigma_i^2}\right] 
\end{equation}
and $\sigma_{i}$ is defined as:
\begin{equation}
    \sigma_i = \sqrt{(\sigma_i^{\rm obs})^2 + (f_i^{\rm obs}\,\sigma_{\rm floor})^2}
\end{equation}
where $\sigma_i^{\rm obs}$ corresponds to the observed flux density uncertainty of the i-eth radio-star, and $\sigma_{\rm floor}\leq0.1$ is an additional fractional variance term, used to ensure posterior smoothness and acceptance fraction $\gtrsim0.1$. Furthermore, the $\sigma_{\rm floor}$ factor can also be interpreted as encoding the intrinsic variability of radio-stars \citep{Cano-Gonzalez2024, Cano-Gonzalez2025}.

The total flux likelihood for a single realisation is then
\begin{equation}
\mathcal{L}_S
=
\prod_{i=1}^{N_{\rm obs}}
 p(f_i^{\rm obs} \mid \boldsymbol{\theta}).
\end{equation}

Because the forward model is stochastic, we averaged over $K$
independent realisations at fixed $\boldsymbol{\theta}$. Therefore, 
\begin{equation}
\mathcal{L}_S^{\rm K}
=
\frac{1}{K}
\sum_{k=1}^{K}
\mathcal{L}_{S,k}.
\end{equation}

We assumed Poisson statistics for the population counts $\mathcal{\rm P} (N|\lambda)=\frac{\lambda^N\, e^{-\lambda}}{N!}$. We estimated the expected values
$\lambda_{\rm WR}$, $\lambda_{\rm OB}$, and $\lambda_{\rm WC}$ by averaging the simulated WN, OB, and WC counts
over the same $K$ realisations used to obtain $\mathcal{L}_S^K$. The Arches cluster contains
$N_{\rm WN}^{\rm obs} = 15$ WN(h) radio-stars,
$N_{\rm OB}^{\rm obs} = 7$ OB radio-stellar systems,
and no detected WC stars ($N_{\rm WC}^{\rm obs} = 0$).

Therefore, the different likelihood terms are
\begin{align}
\mathcal{L}_{\rm WN} &=
\mathrm{P}
\left(
N_{\rm WN}^{\rm obs} \mid \lambda_{\rm WN}
\right), \\[6pt]
 \mathcal{L}_{\rm OB} &=
 \mathrm{P}
\left(
N_{\rm OB}^{\rm obs} \mid \lambda_{\rm OB}
\right), \\[6pt]
 \mathcal{L}_{\rm WC} &=
 \mathrm{P}
\left(
N_{\rm WC}^{\rm obs}  \mid \lambda_{\rm WC}
\right).
\end{align}

In order to ensure numerical stability and avoid convergence issues, we used logarithmic notation. Therefore, the full log-likelihood is expressed as the sum of the flux and population terms:
\begin{equation}
\ln \mathcal{L}(\boldsymbol{\theta})
=
\ln \mathcal{L}_S^{\rm K}
+
\ln \mathcal{L}_{\rm WN}
+
\ln \mathcal{L}_{\rm OB}
+
\ln \mathcal{L}_{\rm WC}.
\end{equation}
where $\ln$ denotes the natural logarithm.

\subsubsection{Priors and posterior}
We considered using a Gaussian prior on $\alpha_{\rm IMF}$, from \citet{Hosek2019},
\begin{equation}
p(\alpha_{\rm IMF}) =
\mathcal{N}(\alpha_{\rm IMF} \mid -1.8,\, 0.11),
\end{equation}
where the uncertainty takes into account both the statistical and systematic uncertainties from \citet{Hosek2019}.
Finally, the posterior distribution is given by Bayes' theorem:
\begin{equation}
\ln \mathcal{P}(\boldsymbol{\theta} \mid \mathrm{data})
=
\ln \mathcal{L}(\boldsymbol{\theta})
+
\ln p(\boldsymbol{\theta})
+
\mathrm{cnst}. 
\end{equation}

\subsubsection{MCMC setup}\label{subsubsect_MCMCsetup}
We used the {\tt emcee} Python package \citep{ForemanMackey2013} to sample the posterior distribution. We used a total of 32 walkers\footnote{In the framework of MCMC, a walker is an individual sampling agent that moves across parameter space according to algorithms such as the Metropolis-Hastings or Gibbs algorithms \citep[e.g.~][]{Hastings1970, Geman1984}.} to explore parameter space, and a total of $2\,500$ iterations. In order to find a suitable compromise between reliable estimates of  expected counts ($\lambda_{\rm WN/OB/WC}$), convergence, and computation time, we set $K=10$ (the number of realisations at fixed $\boldsymbol{\theta}$). The allowed parameter ranges were: $1.5<t_{\rm age}/{\rm Myr}<3.5$, $10\,000<M_{\rm cl}/M_\odot<50\,000$, and $-2.3<\alpha_{\rm IMF}<-1.4$ for cluster age, mass, and IMF slope respectively. We discarded the initial $20\%$ of iterations (burn in) to ensure that the posterior sampling is not biased by initial walker positions. We allowed the IMF to sample stellar masses from $M_{\rm ini}^{\rm min} = 0.1\, M_\odot$ to $M_{\rm ini}^{\rm max}=130\, M_\odot$ \citep{Figer2005}. We note that increasing the maximum value above $130\, M_\odot$ resulted in many unphysical, large thermal fluxes $(\gtrsim1$ mJy, especially in the GENEC models) presumably arising from the mass-loss prescription dependence on luminosity (for these very massive synthetic stars: $\log(L/L_\odot)\gtrsim6.3$). 

Finally, we used DBSCAN \citep{Pedregosa2011} to detect clusters of points within the posterior distributions. The number of samples for a point to be considered as a core point, \texttt{min\_size} was set to ${200-600}$, and the maximum neighbouring distance between two samples, \texttt{eps} was set to ${0.4-0.65}$. The aim behind the previous parameter ranges was to minimize noise in the final cluster (ensuring $\gtrsim90\%$ of posterior points were clustered), and to visually check proper cluster identification.

\subsection{Mock recovery tests}
We performed mock recovery tests in order to estimate the systematic uncertainties in cluster age. For each evolutionary model and metallicity, we quantified age bias as the mean difference between recovered age and true input age $(\Delta t_{\rm age}=t_{\rm age} - t_{\rm age}^{\rm true})$, over 10 MCMC runs, and then we added or subtracted the mean bias to the median value of the posterior distribution. We used three values of $t_{\rm age}^{\rm true}$ for each model, following $(t_{\rm age,1}^{\rm true},t_{\rm age,2}^{\rm true},t_{\rm age,3}^{\rm true})(\rm Myr) = (t_{\rm age}^{\rm obs}-0.5\, ,t_{\rm age}^{\rm obs},\, t_{\rm age}^{\rm obs}+0.5)$, where $t_{\rm age}^{\rm obs}$ is an age close to the  posterior peak returned by the model at a given metallicity. We performed a linear fit in $\Delta t_{\rm age}$-$t_{\rm age}^{\rm true}$ space, and evaluated the polynomial on the observed posterior peak $t_{\rm age}^{\rm obs}$ to estimate the bias on cluster age. Finally, we used $\sigma_{\rm sys}=\sqrt{(\sigma_{\rm 1}^2 + \sigma_{\rm 2}^2 + \sigma_{\rm 3}^2)/3}$, that is, the standard deviation values of $\Delta t_{\rm age}$ in each $t_{\rm age,(123)}^{\rm true}$ to compute the total systematic uncertainty in cluster age.

We refrained from applying the same methodology to $\alpha_{\rm IMF}$ and $M_{\rm cl}$, in the case of flat priors, because the uncertainties of these parameters are dominated by the degeneracy between them. However, when applying \citet{Hosek2019} IMF slope as prior, systematic dispersion in $M_{\rm cl}$ and $\alpha_{\rm IMF}$ becomes comparable to their statistical uncertainties. Therefore, we used these parameters bias dispersion at age posterior peak as a first order assessment of their systematic uncertainties.

In order to make the computational time feasible, during the recovery tests, we lowered the number of total iterations to $1\,000$, as well as reduced the number of realisations per fixed $\boldsymbol{\theta}$ to $K=5$ (we found that $K>3$ was necessary for proper count estimates and convergence), and the number of walkers to 20. Furthermore, as stated in the following section, since cluster age seems uncorrelated with cluster mass and IMF slope, we used fixed $M_{\rm cl}$ and $\alpha_{\rm IMF}$ values of $M_{\rm cl}=25\,000\, M_\odot$ and $\alpha_{\rm IMF}=-1.8$. We also used the IMF value from \citet{Hosek2019} as Gaussian prior on IMF slope for the mock tests, in order to improve convergence. Systematic uncertainties are further discussed in Sect. \ref{subsect_sys}, and shown in Table \ref{Table_posteriorstats}.

\section{Results and discussion}\label{Sect_discussion}
\subsection{Model-averaged estimates}\label{subsect_modelavg}

\begin{figure}
  \centering
  \includegraphics[width=\hsize]{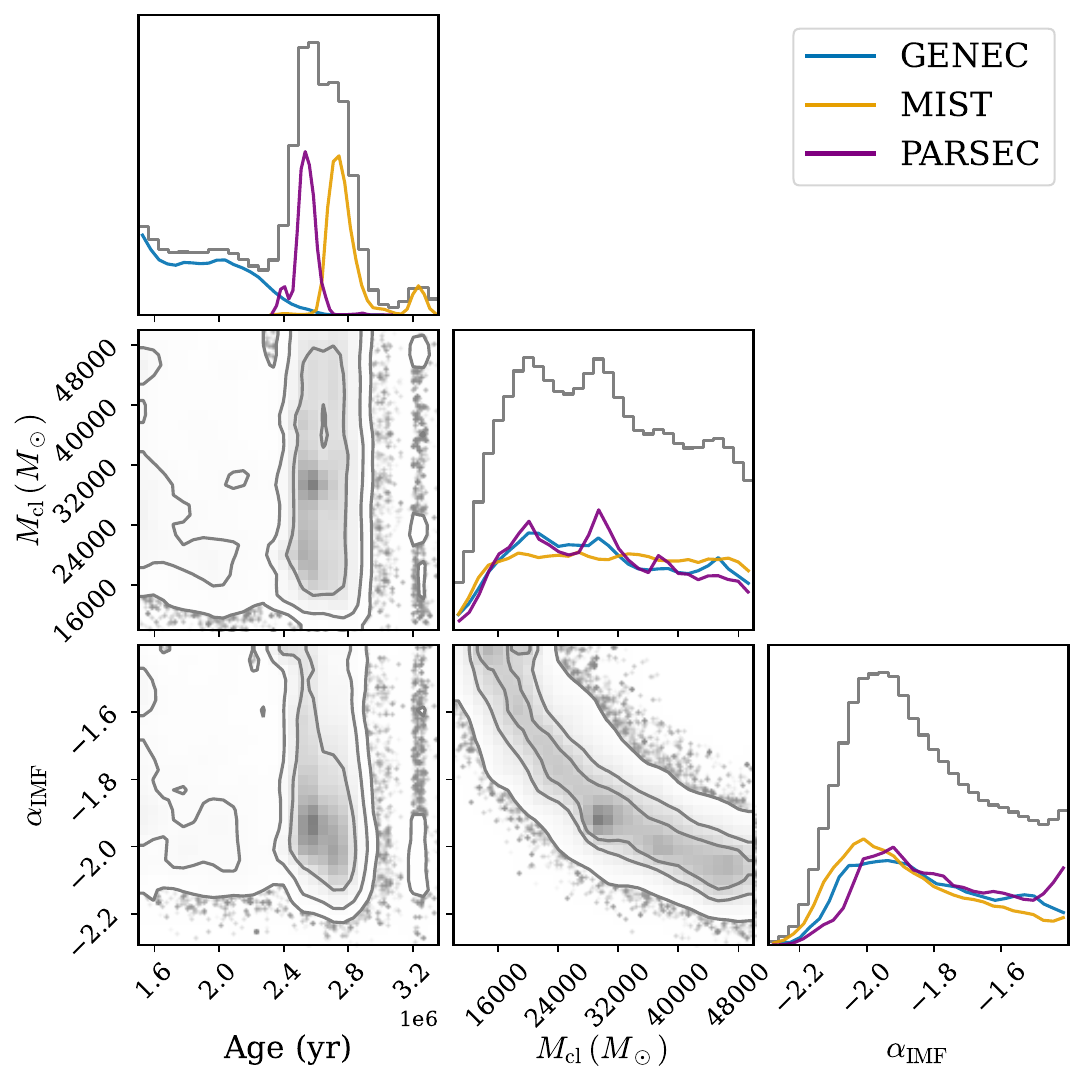}
     \caption{Posterior distribution of the three evolutionary models combined, at solar metallicity. All priors were flat ($\ln p(\boldsymbol{\theta})=0$), which results in a clear degeneracy between $M_{\rm cl}$ and $\alpha_{\rm IMF}$.}
    \label{Fig_CP_flat_Zsol}
\end{figure}

\begin{figure}
  \centering
  \includegraphics[width=\hsize]{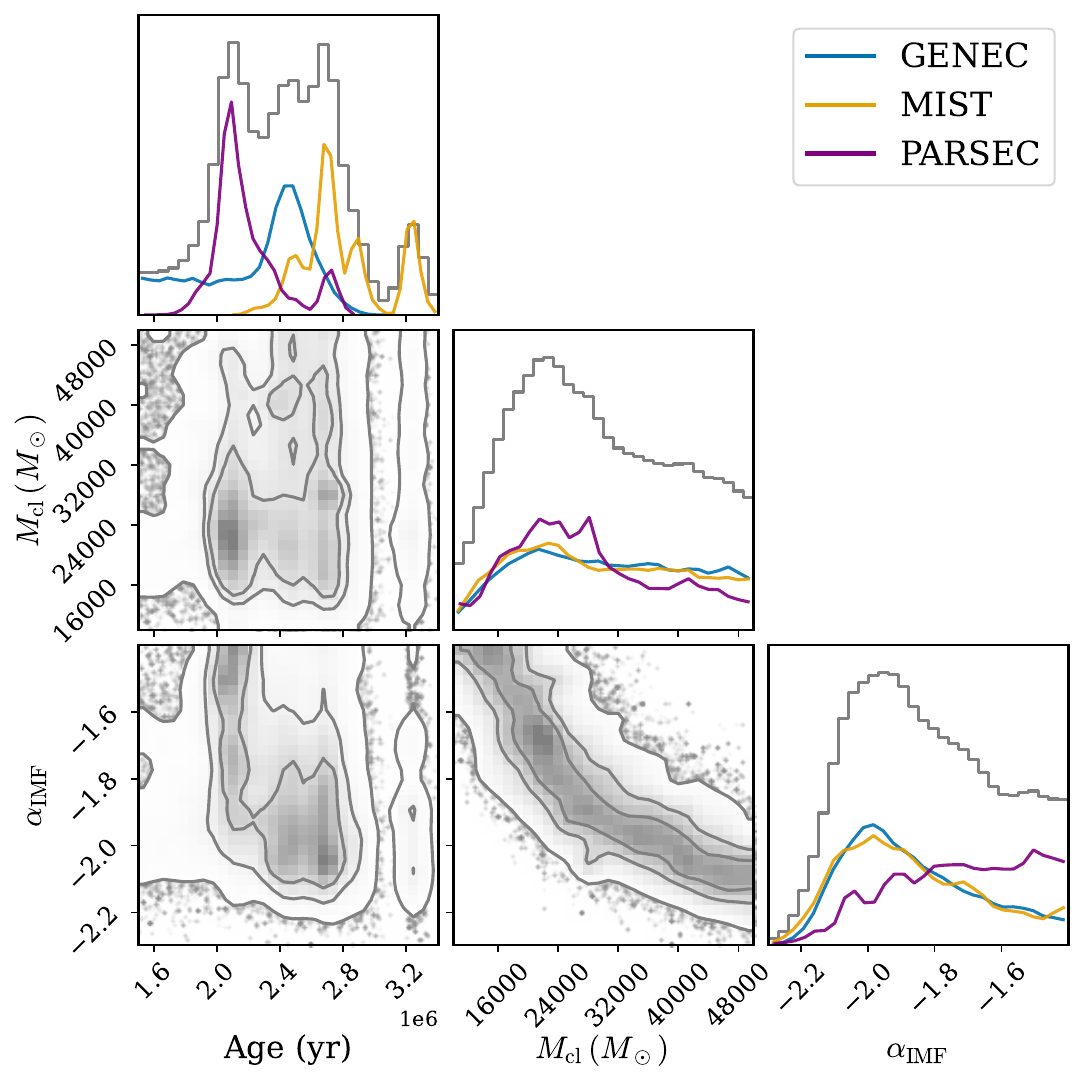}
     \caption{Same as Fig. \ref{Fig_CP_flat_Zsol} but for super-solar metallicity ($Z=0.02$).}
    \label{Fig_CP_flat_Z02}
\end{figure}

\begin{figure}
  \centering
  \includegraphics[width=\hsize]{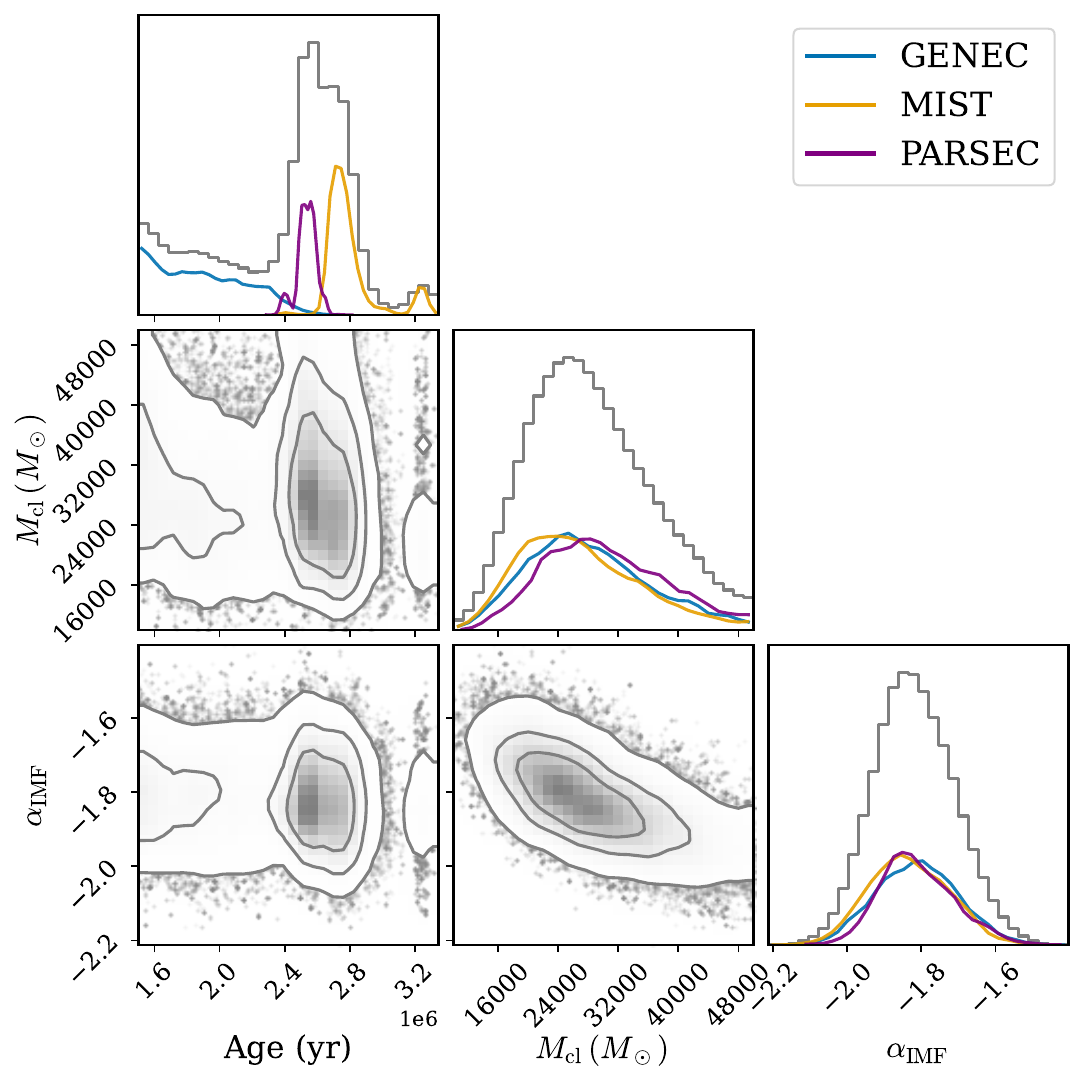}
     \caption{Posterior distribution of the three evolutionary models combined, at solar metallicity, using the IMF slope from \citet{Hosek2019} as model prior.}
    \label{Fig_CP_pH19_Zsol}
\end{figure}

\begin{figure}
  \centering
  \includegraphics[width=\hsize]{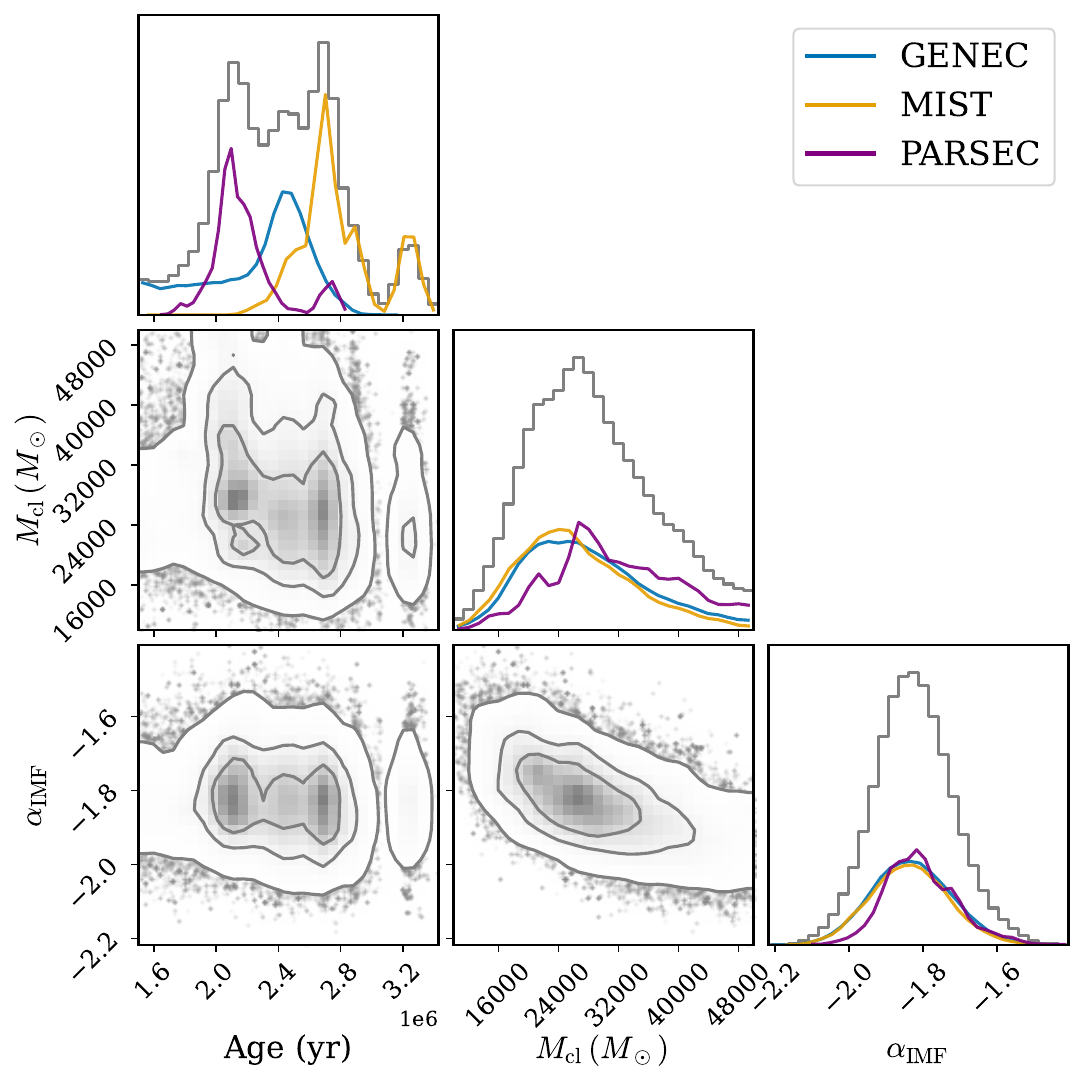}
     \caption{Same as Fig. \ref{Fig_CP_pH19_Zsol}, but at $Z=0.02$.}
    \label{Fig_CP_pH19_Z02}
\end{figure}

Figures \ref{Fig_CP_flat_Zsol} to \ref{Fig_CP_pH19_Z02} show the different posterior distributions returned by the simulations for each metallicity and prior when combining the samples of the three evolutionary models considered. 

Averaging over all three models we obtained preferred ages of $t_{\rm age}=2.54^{+0.22}_{-0.65}$ Myr and $t_{\rm age}=2.43^{+0.32}_{-0.38}$ Myr for $Z=0.014$ and $Z=0.02$, respectively. The previous uncertainties were computed with the median, 16th, and 84th percentiles. These estimations show no significant systematic difference between the considered metallicities. However, when investigating each model separately, we found substantial differences between the solar and super-solar cases in the GENEC and PARSEC models (see, Sect. \ref{subsect_Zdependence}). 

When no priors are used, despite the clear degeneracy between cluster mass and IMF slope (see Figs. \ref{Fig_CP_flat_Zsol} and \ref{Fig_CP_flat_Z02}), our model-averaged results showed quasi-identical preferred IMF slopes for both metallicities. Namely, $\alpha_{\rm IMF}=-1.86^{+0.28}_{-0.18}$ and $\alpha_{\rm IMF}=-1.85^{+0.28}_{-0.20}$ for $Z=0.014$ and $Z=0.02$, respectively, where the values were computed using the median, 16th and 84th percentiles. These results favour a top-heavy IMF, closer in value (and within the uncertainties) to the one obtained by \citet{Hosek2019}. This shows that radio continuum observations, paired with spectroscopic knowledge of the radio-stars under study, can be a powerful, complementary tool to constrain the IMF slope of a YMC.

Motivated by the compatibility between our preferred $\alpha_{\rm IMF}$ values and that provided by \citet{Hosek2019}, we decided to use their IMF slope as model prior in order to better constrain cluster mass, given that the aforementioned $\alpha_{\rm IMF}-M_{\rm cl}$ degeneracy clearly dominates its uncertainties (see Table \ref{Table_posteriorstats}). Thus, we obtained the following cluster mass estimates: $M_{\rm cl}=2.75^{+0.95}_{-0.72}\times10^4\,M_\odot$ and $M_{\rm cl}=2.76^{+1.00}_{-0.72}\times10^4\,M_\odot$ for $Z=0.014$ and $Z=0.02$, respectively. Therefore, there seems to be no overall tendency on cluster mass with metallicity, and we can now establish mass range for the Arches cluster: $20\,000\lesssim  M_{\rm cl}/M_\odot \lesssim 37\,000$.

Individual differences between models and metallicities are explored in Sects. \ref{subsect_diffsim} and \ref{subsect_Zdependence}. Moreover, in Sect. \ref{subsect_lit_comparison}, we compare different literature values to those obtained in this work.

\subsection{Systematic uncertainties}\label{subsect_sys}
Not all recovery tests showed clear, smooth tendencies in age bias with true age. In particular, the GENEC models showed significant degeneracy in recovered age, at around $\gtrsim 2.5$ Myr, independently of $t_{\rm age}^{\rm true}$. This resulted in large, unstable corrections in cluster age, towards $\lesssim2$ Myr. For this reason, we refrained from applying the bias correction and simply added the systematic uncertainty obtained from the bias dispersion.

The MIST models showed a significant underestimation of recovered age, that increases with $t_{\rm age}^{\rm true}$. However, the bias sampled at the three true ages shows significant scatter, and a poor linear fit. Therefore, applying the bias correction resulted in a drastic increase in cluster age, of about $\gtrsim0.6$ Myr, yielding corrected ages of $\gtrsim3.3$ Myr for the Arches cluster. While similar ages have been reported in the literature \citep{Schneider2014,Hosek2019}, we considered this inflated value a product of an statistically unreliable bias correction. Thus, we simply added the rather large systematic errors $(\lesssim0.3\, \rm Myr)$ from the recovery tests to those returned by the percentiles of the posterior distributions.

\begin{figure}
  \centering
  \includegraphics[width=\hsize]{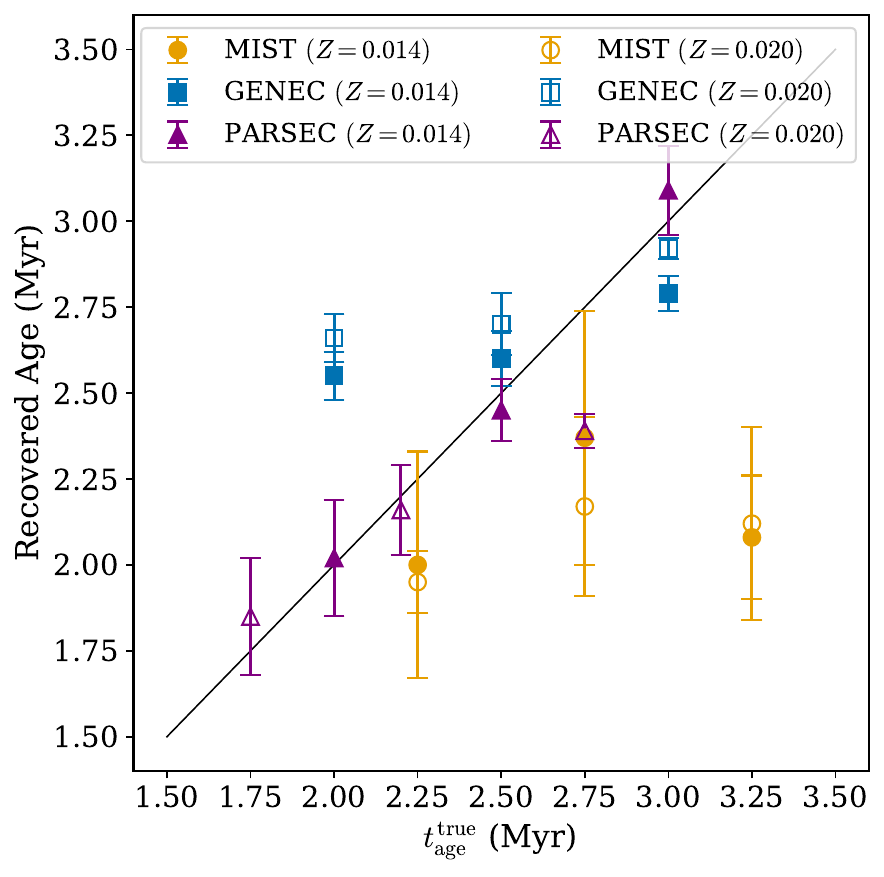}
     \caption{Recovered age versus true input age for all three models at solar (filled symbols) and super-solar (empty symbols) metallicity. The $x=y$ line is shown as measure of goodness of recovery. Note the nearly horizontal tendency of the GENEC models, as well as the large scatter shown by the MIST models. Bias correction is only applied in the case of the PARSEC models.}
    \label{Fig_recage_vs_trueage}
\end{figure}

Finally, the PARSEC models showed insignificant bias, with a moderate slope and relatively small scatter. In this case, resulting in a less drastic correction ($\sim -0.04$ Myr in the case of $Z=0.02$, similar to the statistical uncertainties) to the cluster age. Once the correction is applied, we included the systematic uncertainties derived from the scatter in age bias. Figure\,\ref{Fig_recage_vs_trueage} shows the recovered age versus the true input age for all three evolutionary models.

As mentioned in Sect. \ref{Sect_methods}, systematic uncertainties in $M_{\rm cl}$ and $\alpha_{\rm IMF}$ (for the case in which the IMF prior was applied) were computed with the standard deviation of their respective recovered values, when $t_{\rm age}^{\rm true}$ is evaluated at posterior peak $(t_{\rm age,2}^{\rm true})$. Table \ref{Table_posteriorstats} shows the systematic uncertainties along the values returned by the posterior distributions.

\subsection{Model comparison}\label{subsect_diffsim}
Table \ref{Table_posteriorstats} shows the preferred cluster parameters for all individual evolutionary models and metallicities, including both statistic and systematic uncertainties. 
\def\arraystretch{1.5}
\setlength{\tabcolsep}{10pt}
\begin{table*}
\label{Table_posteriorstats}
 \caption{Statistics extracted from the posterior distributions.}
 \label{Table_posteriorstats}
\centering
\begin{tabular}{lccc}
\hline \hline
Z & $t_{\rm age}$ (Myr) & $M_{\rm cl}\ (\times10^4M_\odot)$  & $\alpha_{\rm IMF}$  \\
\hline
\multicolumn{4}{c}{GENEC} \\
\hline
0.014 & $\lesssim2.5$ & $2.9^{+1.8}_{-1.4}$ $\left(2.72^{+0.92}_{-0.70}\,(\rm stat.)\pm 0.28\,(sys.)\right)$  & $-1.85^{+0.38}_{-0.25}$ $\left(-1.81^{+0.11}_{-0.11}(\rm stat.)\pm 0.02\,(sys.) \right)$ \\
0.020 & $2.38^{+0.18}_{-0.56}$ (stat.) $\pm\,0.07$ (sys.) & $2.9^{+1.8}_{-1.5}$ $\left(2.67^{+0.94}_{-0.69}(\rm stat.)\pm 0.30\,(sys.)\right)$ &  $-1.90^{+0.26}_{-0.16}$ $\left(-1.83^{+0.11}_{-0.10}(\rm stat.)\pm 0.02\,(sys.)\right)$\\
\hline

\multicolumn{4}{c}{MIST} \\

\hline
0.014 & $2.75^{+0.08}_{-0.06}$ (stat.) $\pm\,0.30$ (sys.) & $3.1^{+1.7}_{-1.6}$ $\left(2.61^{+0.91}_{-0.69}(\rm stat.)\pm 0.26\,(sys.)\right)$  & $-1.91^{+0.26}_{-0.16}$ $\left(-1.84^{+0.11}_{-0.10}(\rm stat.)\pm 0.01\,(sys.)\right)$\\
0.020 & $2.69^{+0.17}_{-0.18}$ (stat.) $\pm\,0.22$ (sys.) & $2.9^{+1.9}_{-1.5}$ $\left(2.59^{+0.87}_{-0.65}(\rm stat.)\pm 0.35\,(sys.)\right)$ &  $-1.89^{+0.25}_{-0.17}$ $\left(-1.84^{+0.10}_{-0.10}(\rm stat.)\pm 0.03\,(sys.)\right)$\\
\hline

\multicolumn{4}{c}{PARSEC} \\

\hline
0.014 & $2.54^{+0.05}_{-0.05}$ (stat.) $\pm\,0.14$ (sys.) & $2.9^{+1.7}_{-1.4}$ $\left(2.95^{+0.94}_{-0.73}(\rm stat.)\pm 0.33\,(sys.)\right)$  & $-1.81^{+0.38}_{-0.26}$ $\left(-1.82^{+0.11}_{-0.08}(\rm stat.)\pm 0.02\,(sys.)\right)$\\
0.020 & $2.15^{+0.19}_{-0.08}$ (stat.) $\pm\,0.13$ (sys.) & $2.7^{+1.9}_{-1.2}$ $\left(3.12^{+0.97}_{-0.69}(\rm stat.)\pm 0.27\,(sys.)\right)$ &  $-1.69^{+0.26}_{-0.36}$ $\left(-1.80^{+0.09}_{-0.08}(\rm stat.)\pm 0.02\,(sys.)\right)$\\
\hline\hline
\end{tabular}
\tablefoot{In the case of bi-modality, only the dominant DBSCAN clusters are shown. The statistical uncertainties were computed with the median, 16th, and 84th percentiles. The values between parenthesis represent the results obtained using the prior on IMF slope from \citet{Hosek2019}. When flat priors are used, the $M_{\rm cl}$ and $\alpha_{\rm IMF}$ distributions may not be Gaussian-like, therefore, we show the median plus the 90\% credible intervals, as they better represent the simulated data.}

\end{table*}

For all models and metallicities, we found that the estimated ages are independent from and uncorrelated with cluster mass and IMF slope (see the almost identical age distributions in e.g.~ Figs\, \ref{Fig_CP_flat_Z02} and \ref{Fig_CP_pH19_Z02}), with the corresponding correlation coefficient\footnote{The correlation coefficient, $\eta_c$ was computed using Numpy \texttt{corrcoef} function (\url{https://numpy.org/doc/stable/reference/generated/numpy.corrcoef.html}).} $\eta_c<0.2$ and mostly $\lesssim0.1$. On the contrary,  for $M_{\rm cl}$ and $\alpha_{\rm IMF}$ we obtained $\eta_c\gtrsim0.7$. With flat priors we always obtained a strong degeneracy between $M_{\rm cl}$ and $\alpha_{\rm IMF}$ for all evolutionary models and assumed metallicities.

Contrary to the other two stellar evolutionary codes, the PARSEC models (in the 1.2S release) do not take rotation into account. The lack of stellar rotation shortens stellar lifetime \citep[see][Table 1]{Georgy2012}. This effect can be seen in the bottom panel of Fig. \ref{Fig_WRcounts}. In PARSEC, stars evolve more rapidly along the different WR stages (schematically: ${\rm WN}\rightarrow{\rm WC}\rightarrow {\rm WO}$), resulting in the appearance of WC stars at $\lesssim 2.9$ Myr. This will naturally impact the $\mathcal{L}_{\rm WC}$ component of the likelihood, making a WC-rich solution less likely to be chosen by the MCMC sampler. This is possibly the reason why PARSEC models mostly return $t_{\rm age}\lesssim2.5$ Myr. Unfortunately, at the time of writing, the PARSEC V2 release, which includes stellar rotation, only covers up to 14 $M_\odot$ \citep{Nguyen2025}, making it unsuitable for the work presented here.

\subsection{Metallicity dependence}\label{subsect_Zdependence}
GENEC models, at first glance, show different preferred $t_{\rm age}$ posterior distributions between the considered metallicities. Namely, the solar metallicity model does not return a Gaussian-like age posterior, but rather an upper limit at $\lesssim2.5\, {\rm Myr}$, preceded by a continuous distribution towards younger ages, with no discernable spikes or over-densities (see Fig. \ref{Fig_CP_flat_Zsol}). 
On the contrary, the GENEC models at super-solar metallicity ($Z=0.02$) returned a more morphologically defined posterior distribution, peaking at $\approx2.4\ {\rm Myr}$, (see Fig. \ref{Fig_CP_flat_Z02}). However, the continuous tail towards younger ages is still present, albeit less pronounced. All this resulted in a rather large lower statistical uncertainty: $t_{\rm age}=2.38^{+0.18}_{-0.56}$ Myr. We were not able to find an explanation for the morphological difference between the two age posterior distributions at the two considered metallicities. We found that, for both metallicities, average mass-loss rates of synthetic WN stars increased as a function of age, with the super-solar models returning slightly higher mass-loss rates on average (with a mean difference of $\dot{M}_{\rm Z02} - \dot{M}_{\rm Z_\odot}\lesssim10^{-5}\,M_\odot\,\rm yr^{-1}$). However, at all ages tested, the values obtained at both metallicities were consistent with the mass-loss values reported in \citet{Cano-Gonzalez2024} for the Arches WNh stars.

We could also see a clear age-metallicity dependence using PARSEC models, with the higher metallicity resulting in lower ages (see Figs. \ref{Fig_CP_flat_Zsol} and \ref{Fig_CP_flat_Z02}). One of the reasons for this behaviour may be the fact that PARSEC seems to be the model in which WN counts are most affected by metallicity (see, Fig. \ref{Fig_WRcounts}). Namely, higher metallicities systematically returned more synthetic WN stars in the 2-2.5 Myr range. Moreover, it is worth mentioning that PARSEC models returned the lowest mean acceptance fraction\footnote{The acceptance fraction of an MCMC sampler is the ratio of accepted moves to the total number of moves. It is computed per walker.} during the MCMC simulations (as low as $8\%$), which is not catastrophic, but indicates that these simulations struggled more with finding stable regions in parameter space. 

Finally, we could not find any significant differences between the considered metallicities in the MIST models. Both the solar and super-solar simulations returned closely similar values for all explored parameters. Furthermore, both metallicities returned bi-modal distributions sitting at roughly the same points of parameter space.

\subsection{Bi-modality and clustering}\label{subsect_clustering}
We found bi-modality in the age posterior distributions of some of the evolutionary models, particularly MIST and PARSEC. In all cases there is a clearly dominant cluster of points in parameter space. However, using the full posterior statistics in these cases would surely misrepresent the underlying results. For example, we can clearly appreciate two over-densities around $\sim2.75\, \rm Myr$ and $\sim3.25\, \rm Myr$ in the yellow histogram (MIST) of Fig. \ref{Fig_CP_flat_Z02}. On the other hand, PARSEC models at $Z=0.02$ also show bi-modality in their age distribution, with over-densities located at $\sim2.1$ Myr and $\sim2.7$ Myr. Table \ref{Table_secondarystats} shows the statistics extracted from the secondary clusters detected by DBSCAN.

\subsection{Contribution of each likelihood term}
In order to study the different contributions of $\mathcal{L}_S^K$ and $\mathcal{L}_{\rm WN/OB/WC}$ over the final posterior distribution, we run analogous simulations with only one of these likelihood terms passed to the MCMC sampler. We found that posterior $t_{\rm age}$ distributions are mostly modulated by the $\mathcal{L}_S^{K}$ term, while the $\mathcal{L}_{\rm WN/OB/WC}$ terms encode the degeneracy between $M_{\rm cl}$ and $\alpha_{\rm IMF}$, and constrain the upper limit on cluster age. These tendencies were particularly clear with the GENEC and MIST models, while the $\mathcal{L}_{\rm WN/OB/WC}$ term in PARSEC seems to establish a more restrictive upper limit in age, possibly due to the early appearance of WC stars. Appendix \ref{App_Lterms} shows different corner plots for all the evolutionary models, but using only one component of the likelihood. 

Given that the distribution of the observed stellar radio flux densities provides a well defined posterior distribution for cluster age, we conclude that the measured radio flux densities of stars can be a highly useful tool to constrain cluster ages without any previous spectroscopic information on stellar type. This is ideal if our aim is to constrain the age of diluted clusters (co-moving groups of massive stars, \citealp{Martinez-Arranz2024}) that possibly lie within the GC but have not been identified yet because of intense stellar crowding, reddening, and limited spectroscopic campaigns.

\subsection{Comparison with previous literature}\label{subsect_lit_comparison}

\begin{figure}
  \centering
  \includegraphics[width=\hsize]{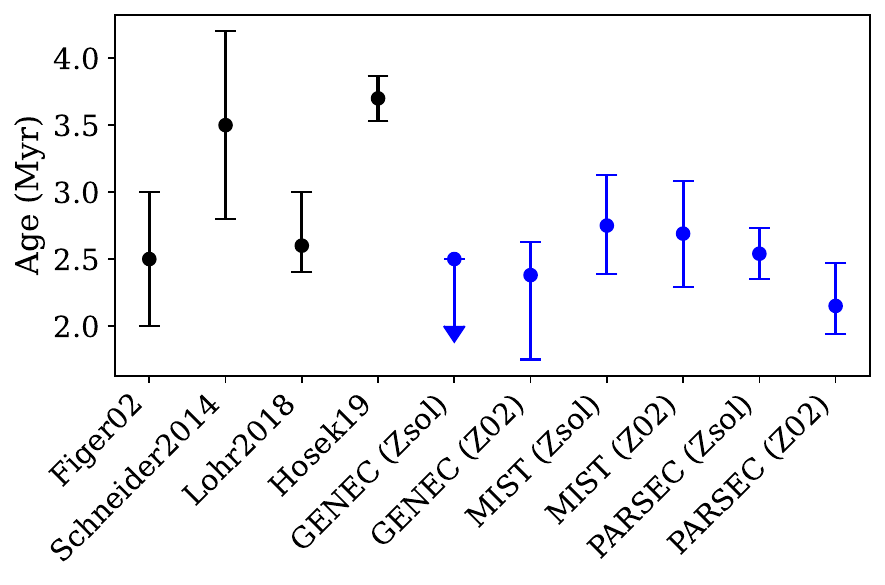}
     \caption{Comparison of estimated ages of the Arches cluster in previous literature (black) and our results (blue).
     From left to right, the values are obtained from \citet{Figer2002,Schneider2014,Lohr2018,Hosek2019}.}
    \label{Fig_age_comparison}
\end{figure}

Previous estimates of the Arches cluster age can be grouped into two broad sets: estimates centred around a younger age of $\sim2.5$ Myr \citep{Figer2002,Lohr2018,Clark_IV}, and somewhat older estimates centred around $\sim 3.5$ Myr \citep{Schneider2014,Hosek2019}. Figure \ref{Fig_age_comparison} illustrates the comparison between all our age estimates and values reported in previous literature. As shown in Sect. \ref{subsect_modelavg}, our results are clearly more compatible with the former group. However, the MIST models are still marginally consistent with the value provided by \citet{Schneider2014}, mainly because of the large systematic uncertainties of MIST age estimates. The age provided by \citet{Hosek2019} is clearly incompatible with our estimates. We theorise that this older value could come from the WR stage definition of the evolutionary models used in their work (MIST v1.0), which, as mentioned in Sect. \ref{subsect_thermalemission}, do not include hydrogen-rich ($X_{\rm H} \sim0.5$) WR stars. Perhaps, if this set of less hydrogen-depleted WR stars had been used in the simulations of \citet{Hosek2019}, their models would have favoured younger ages, maybe even compatible to those derived in this work. Nevertheless, the secondary clusters shown in Table \ref{Table_secondarystats} are more consistent with the older group of age estimates, however, these secondary DBSCAN clusters are clearly not dominant and are not present in all evolutionary models or metallicities.

\begin{figure}
  \centering
  \includegraphics[width=\hsize]{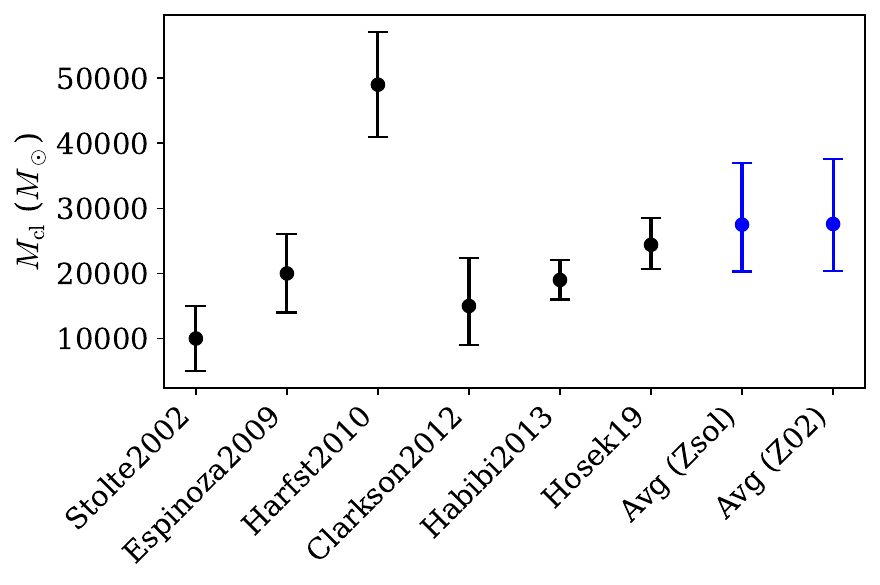}
     \caption{Comparison of estimated masses of the Arches cluster in previous literature (black) and our model-averaged results (blue). Chronologically, from left to right, mass values from \citet{Stolte2002,Espinoza2009,Harfst2010, Clarkson2012,Habibi2013, Hosek2019}.}
    \label{Fig_mass_comparison}
\end{figure}

Mass estimates of the Arches cluster span a wide range of values depending on methodology, assumptions on IMF, etc. Figure \ref{Fig_mass_comparison} shows a comparison between our mass estimates and some values reported in the literature. Initial cluster masses of $\gtrsim50\,000\,M_\odot$ are usually derived under the assumption of a Salpeter-like IMF \citep[e.g.~][]{Harfst2010}, which seems in disagreement with our results. Surely, fixing the IMF slope to $\alpha_{\rm IMF}\sim-2.3$ and increasing cluster mass above the $M_{\rm cl}\sim50\,000\, M_\odot$ range, would produce a similar number of synthetic radio-stars in our simulations, potentially resulting in analogous results to the ones obtained under the assumption of a top-heavy IMF, and smaller cluster masses. On the other hand, some studies focus on the present-day mass function of the cluster by analysing the infrared counts of the inner core $(\lesssim1\, \rm pc)$ of Arches \citep[e.g.~][]{Stolte2002,Espinoza2009,Clarkson2012,Habibi2013}. With this framework, the aforementioned works derive cluster masses of $\lesssim20\,000\,M_\odot$ (we assumed a $50\%$ relative error in the value reported by \citealt{Stolte2002}). We obtained larger, yet compatible cluster mass values, whose difference we attribute to the fact that we are deriving total initial cluster mass, rather than the present day cluster mass of the cluster core, which is bound to be smaller due to tidal stripping and mass-loss of the most massive stars. Finally, our mass estimates are in good agreement with that of \citet{Hosek2019}. This is expected, as our best results on cluster mass are obtained using their IMF slope value as prior. 

As mentioned in Sect. \ref{subsect_modelavg}, our estimates on IMF slope are similar and compatible ($\sim0.2\sigma$) with the one provided by \citet{Hosek2019}. If we assume a Salpeter IMF slope of the form $\alpha_{\rm IMF}^S=-2.35\pm0.2$, we find that our results are $\sim1.6\sigma$ compatible with this value. While we cannot claim significant statistical incompatibility with the Salpeter IMF, it is clear that a top-heavy IMF is preferred by our simulations. Unfortunately, many of the previous works on the Arches mass-function focus on the present-day mass function, based on observations of the cluster's projected core \citep{Stolte2002,Kim2006,Espinoza2009,Habibi2013}. Since our simulations do not include spatial distribution of synthetic radio-stars, it is misleading to directly compare their present-day mass function values to our IMF estimates. The only exception within the previous works is that of \citet{Kim2006}, which estimates an IMF slope of $\alpha_{\rm IMF}^{\rm K06}=-2.05\pm0.05$, that is $\sim0.8\,\sigma$ compatible with our results.

\subsection{Expanding the method towards other YMCs}
Applying this method to other YMCs requires stating some other implicit assumptions inherent to this work, in addition to those mentioned in section\,\ref{subsect_initassump}. Namely, we need to  detect all radio stars individually, which limits this method to Galactic clusters with current instruments. We also assume that the target cluster formed during a coeval event. Multiple episodes of star formation can be modelled by adding two clusters formed at different ages, however, this would require expanding the parameter space with the mass created in each star forming event, the timespan between them, and their respective IMF slopes. Finally, we also need accurate spectral indices for the radio stars. 
 
The Quintuplet cluster could in principle be subject of an analogous study, however, some Arches-like WNh stars are found within its massive stellar population \citep{Clark2018_Quint}, pointing towards some degree of non-coevality, and many of its radio-stars present rather ambiguous radio emission type, denoted by large spectral index uncertainties \citep{Cano-Gonzalez2025}. Furthermore, 8/41 of the Quintuplet radio-stars reported in \citet{Cano-Gonzalez2025} have not been spectroscopically identified yet, making the initial assumptions shown in Sect. \ref{subsect_initassump} less reliable for this YMC.

Another inevitable source of systematic uncertainty is the fact that most massive stars will interact with a companion throughout their lifetimes \citep{Sana2012}, and that these binary interactions can play a significant role in the evolution of massive stars via a myriad of physical processes \citep[e.g.~][for a recent review]{Marchant2024_review}. In the case of the Arches cluster, we adopted the notion proposed in \citet{Clark_IV}, that the cluster is too young for the multiple systems to have reached a relaxed state, indicated by the short-period, high-eccentricity orbits analysed in their work. Within this framework, we can assume that different massive members of a potentially multiple system evolve separately. Thus, the use of single stellar evolutionary models is justified.

\begin{figure}
  \centering
  \includegraphics[width=\hsize]{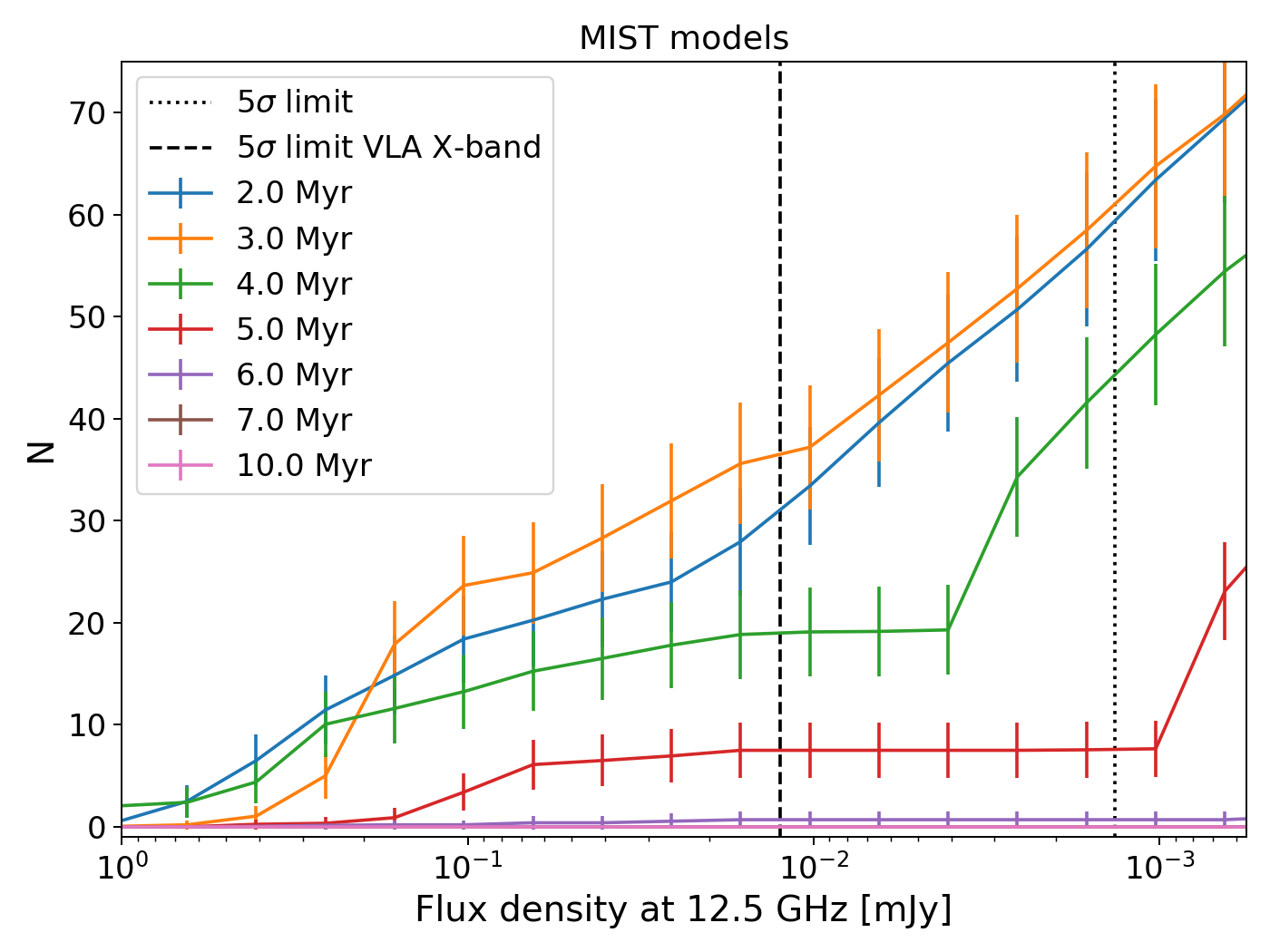}
     \caption{MIST radio luminosity function of a $20\,000\, M_\odot$ cluster for different ages. Vertical lines represent the $5\sigma$ sensitivity limit achieved with the VLA in (dashed line, \citealp{Cano-Gonzalez2024}) and the SKAO-Mid sensitivity assuming 10 h of observations at band 5b (dotted line). Figure credits: \citet{SKA_whitebook}. }
    \label{Fig_RLF}
\end{figure}

In the coming years, the \textit{Square Kilometre Array Observatory} (SKAO) will push the boundaries of radio studies in the GC, in particular at cm-wavelengths (SKA-Mid). With its privileged position in the southern hemisphere, and improved resolution and sensitivity, we will be able to sample the Arches FDD towards fainter flux densities, helping test and refine the hypotheses presented in this work. Figure \ref{Fig_RLF} shows the radio luminosity function derived from MIST models for different cluster ages representing a starburst event of $20\,000\, M_\odot$. We can see in Fig. \ref{Fig_RLF} that the ages decouple better towards fainter flux densities, which will make SKA-Mid a useful tool to estimate YMC ages. 

Furthermore, other Galactic YMCs, such as Westerlund 1 will be observed in cm radio wavelengths with sub-arcsecond resolution for the first time with SKA-Mid, providing deeper insight into key properties of its massive stellar population, and complementing studies carried out at different wavelengths \citep[e.g.~][]{Guarcello2024}.

\section{Summary and conclusions}\label{Sect_conclusions}
We use radio continuum data of massive stars in the Arches cluster to constrain three key parameters: cluster age, cluster mass, and IMF slope.

We perform different sets of MCMC simulations for three evolutionary models: GENEC, MIST, and PARSEC, and for two different metallicities: solar ($Z=0.014$) and super-solar ($Z=0.02$). We build a two-term likelihood. The first term compares flux densities from synthetic clusters to the observations, and the second term compares counts of stellar types relevant to the Arches cluster, namely, WN stars as tracers of thermal radio-stars, OB systems as tracers of non-thermal, colliding-wind binaries, and WC stars.  

We find that cluster age is largely uncorrelated with both cluster mass and IMF slope. We also find that the age posterior distribution is mostly dictated by the flux term of the likelihood, while the expected degeneracy between cluster mass and IMF slope is encoded in the counts-term of the likelihood. This is particularly relevant if we wish to study newly discovered clusters whose massive stellar population has not been spectroscopically characterised yet. Precisely, the GC is thought to harbour many unknown clusters, cloaked by the extreme crowding and reddening.

We conclude that the Arches radio flux density distribution carries substantial age information and favours a young age, around 2.5 Myr, averaging over the evolutionary models considered. We attribute individual differences between models to the differences in the implementation of the models' physical ingredients, especially mass-loss rates and stellar rotation. Nevertheless, regardless of the evolutionary model, our results are compatible with ages reported in \citet{Figer2002, Lohr2018, Clark_IV}.

We find that the preferred cluster mass is, when the IMF prior from \citet{Hosek2019}
is adopted, around $2.7 \times 10^4 M_\odot$. We constrain the Arches cluster mass within $20\,000\lesssim M_{\rm cl}/M_\odot\lesssim37\,000$.

Our radio data favour a top-heavy IMF, consistent with \citet{Hosek2019}. However, this is not yet a decisive, statistical exclusion of a Salpeter IMF, mainly because of the cluster
mass-IMF degeneracy.

Radio continuum observations can become a useful, complementary tool for YMC parameter
inference. However, a general spectroscopy-free method will require
better calibration of the radio luminosity functions of (thermal) WR stars and (non-thermal) colliding-wind binaries. Future radio interferometers like SKA-Mid, will provide a quality jump in GC radio-continuum observations, allowing us to refine the methodology presented in this work.

\begin{acknowledgements}

We warmly thank the NRAO staff for their guidance during the data reduction process that made this work possible.

MCG, RS, AA, JM, F.N.-L, and MPT acknowledge financial support from the Severo Ochoa grant CEX2021-001131-S funded by MCIN/AEI/ 10.13039/501100011033. MGC and RS acknowledge support from grant EUR2022-134031 funded by MCIN/AEI/10.13039/501100011033 and by the European Union NextGenerationEU/PRTR and by grant PID2022- 136640NB-C21 funded by MCIN/AEI 10.13039/501100011033 and by the European Union.
Authors AA, JM and MPT acknowledge financial support from the Spanish grant PID2023-147883NB-C21, funded by MCIU/AEI/ 10.13039/501100011033, as well as support through ERDF/EU.

AA acknowledges support by the PID2023-147883NB-C21 grant.

F.N., acknowledges support by grant PID2022-137779OB-C41 funded by
MCIN/AEI/10.13039/501100011033 by "ERDF A way of making Europe".

Authors MCG, RS and JM acknowledge the Spanish Prototype of an SRC (SPSRC) service and support funded by the Ministerio de Ciencia, Innovación y Universidades (MICIU), by the Junta de Andalucía, by the European Regional Development Funds (ERDF) and by the European Union NextGenerationEU/PRTR. The SPSRC acknowledges financial support from the Agencia Estatal de Investigación (AEI) through the "Center of Excellence Severo Ochoa" award to the Instituto de Astrofísica de Andalucía (IAA-CSIC) (SEV-2017-0709) and from the grant CEX2021-001131-S funded by MICIU/AEI/ 10.13039/501100011033.

JM acknowledges financial support from the grant  PID2021-123930OB-C21 funded by MICIU/AEI/ 10.13039/501100011033 and by ERDF/EU.

F.N.-L. gratefully acknowledges financial support from the Ramón y Cajal programme (RYC2023-044924-I), funded by MCIN/AEI/10.13039/501100011033 and FSE+; from grant PID2024-162148NA-I00, funded by MCIN/AEI/10.13039/501100011033 and the European Regional Development Fund (ERDF) “A way of making Europe”.
\end{acknowledgements}

\bibliographystyle{aa} 
\bibliography{references.bib} 

\begin{appendix}

\section{Secondary DBSCAN cluster statistics}

\def\arraystretch{1.3}
\setlength{\tabcolsep}{7pt}
\begin{table*}\label{Table_secondarystats}
 \caption{Statistics extracted from the secondary DBSCAN clusters.}
 \label{Table_secondarystats}
\centering
\begin{tabular}{lccc}
\hline \hline
Z & $t_{\rm age}$ (Myr) & $M_{\rm cl}\ (\times10^4M_\odot)$  & $\alpha_{\rm IMF}$  \\
\hline
\multicolumn{4}{c}{MIST} \\
\hline
0.014 $^{(0.1)}$ & $3.23^{+0.02}_{-0.03}$ (stat.) $\pm$ 0.30 (sys.) & $2.7^{+2.1}_{-1.3}$ $\left(2.29^{+1.12}_{-0.44}(\rm stat.)\pm 0.26\,(sys.)\right)$  & $-1.91^{+0.39}_{-0.24}$ $\left(-1.84^{+0.11}_{-0.08}(\rm stat.)\pm 0.01\,(sys.)\right)$\\
0.020 $^{(0.2)}$& $3.24^{+0.03}_{-0.03}$ (stat.) $\pm$ 0.22 (sys.) & $2.7^{+2.0}_{-1.4}$ $\left(2.35^{+0.86}_{-0.57}(\rm stat.)\pm 0.35\,(sys.)\right)$ &  $-1.91^{+0.25}_{-0.18}$ $\left(-1.83^{+0.10}_{-0.09}(\rm stat.)\pm 0.03\,(sys.)\right)$\\
\hline
\multicolumn{4}{c}{PARSEC} \\
\hline
0.020 $^{(0.07)}$ & $3.07^{+0.04}_{-0.06}$ (stat.) $\pm$ 0.13 (sys.) & $2.4^{+0.5}_{-1.2}$ $\left(2.7^{+1.3}_{-1.0}(\rm stat.)\pm 0.3\,(sys.)\right)$ &  $-1.90^{+0.36}_{-0.18}$ $\left(-1.82^{+0.02}_{-0.08}(\rm stat.)\pm 0.02\,(sys.)\right)$\\
\hline\hline
\end{tabular}
\tablefoot{Same statistics as in Table \ref{Table_posteriorstats}. The value in parenthesis next to the metallicity represents the fraction of posterior data that corresponds to the secondary cluster.}

\end{table*}

\section{Counts-only and flux-only corner plots}\label{App_Lterms}

\begin{figure}
  \centering
  \includegraphics[width=\hsize]{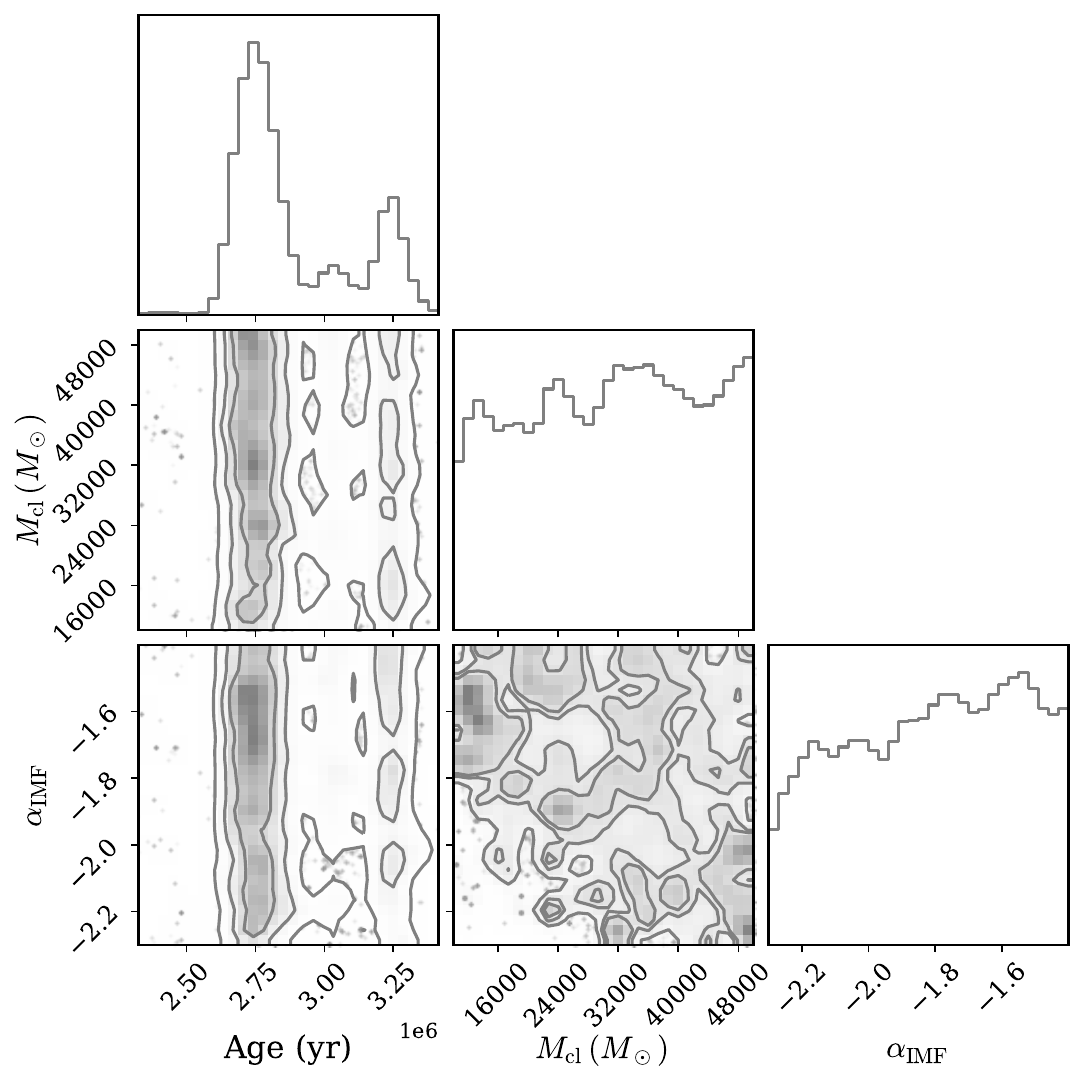}
     \caption{MIST posterior distributions at solar metallicity using only the flux term of the likelihood. Note the similarity between the age distribution here and the one shown in the blue histogram of Fig. \ref{Fig_CP_flat_Zsol}, which uses the full likelihood.}
    \label{Fig_fluxonly_MIST}
\end{figure}

\begin{figure}
  \centering
  \includegraphics[width=\hsize]{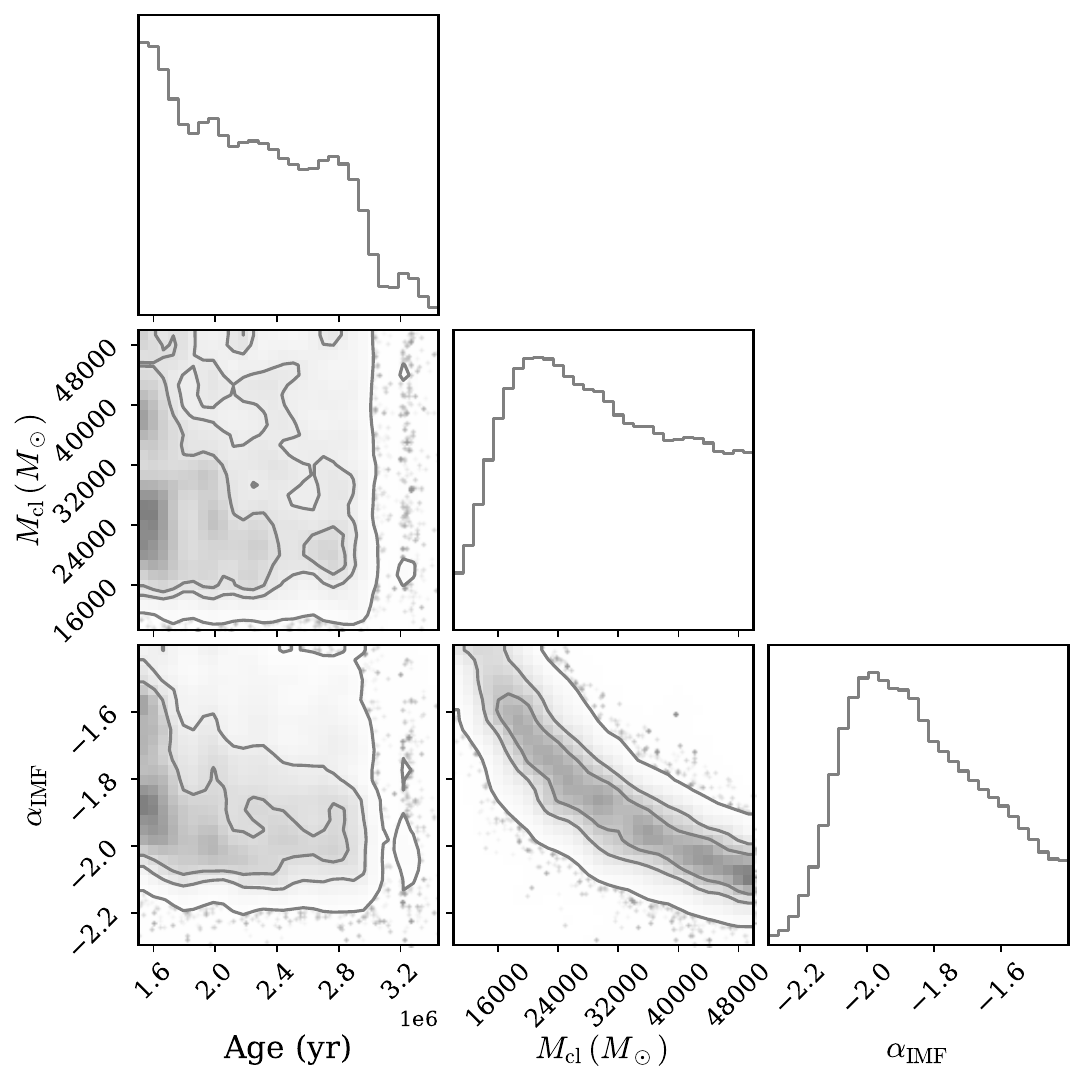}
     \caption{MIST posterior distributions at solar metallicity for a counts-only likelihood. Note the well-defined degeneracy between IMF slope and cluster mass, as well as the wide age range covered in this case, with no clearly defined over-densities or peaks.}
    \label{Fig_countsonly_MIST}
\end{figure}

\begin{figure}
  \centering
  \includegraphics[width=\hsize]{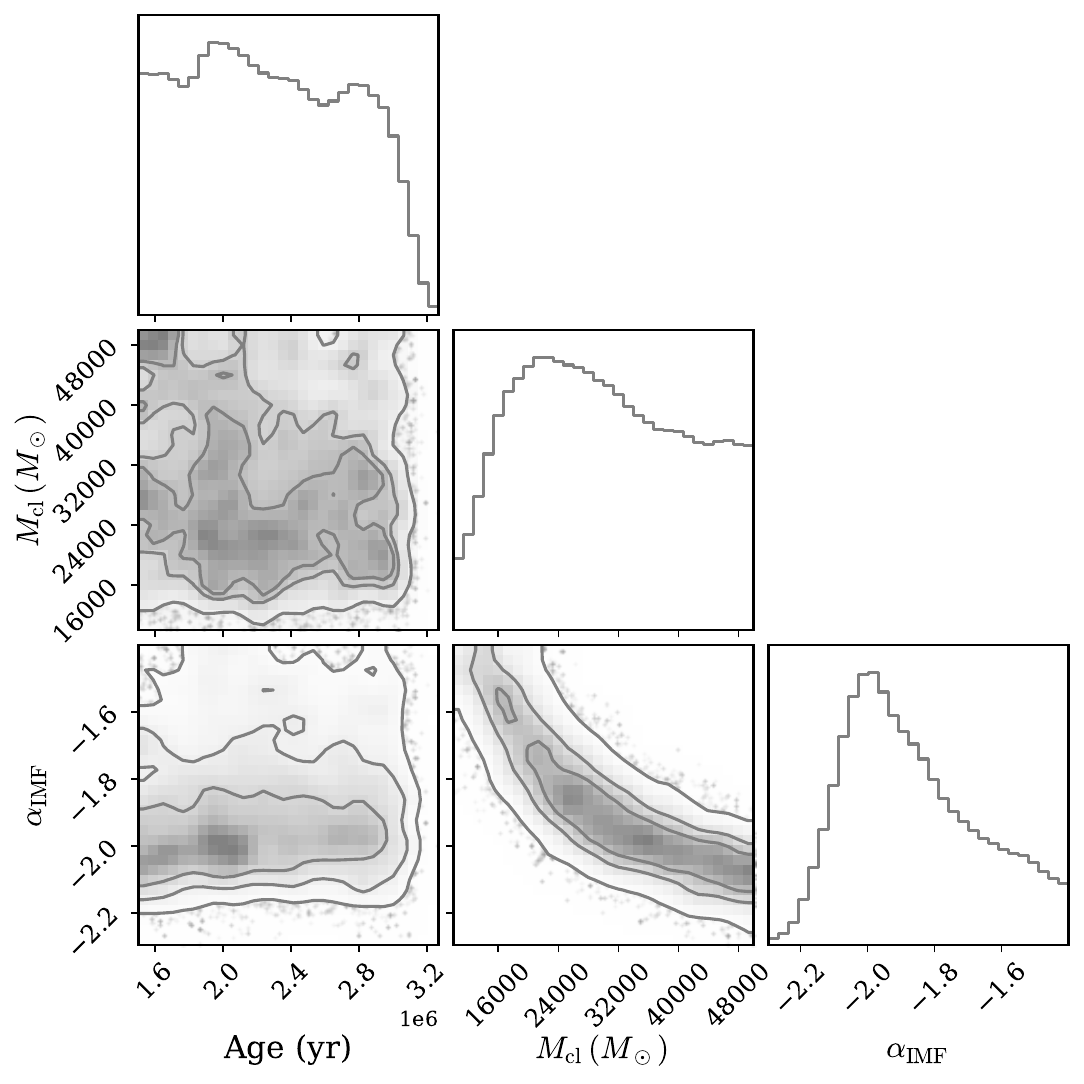}
     \caption{GENEC posterior distributions at $Z=0.02$ for a counts-only likelihood. }
    \label{Fig_countsonly_Ekstrom}
\end{figure}

\begin{figure}
  \centering
  \includegraphics[width=\hsize]{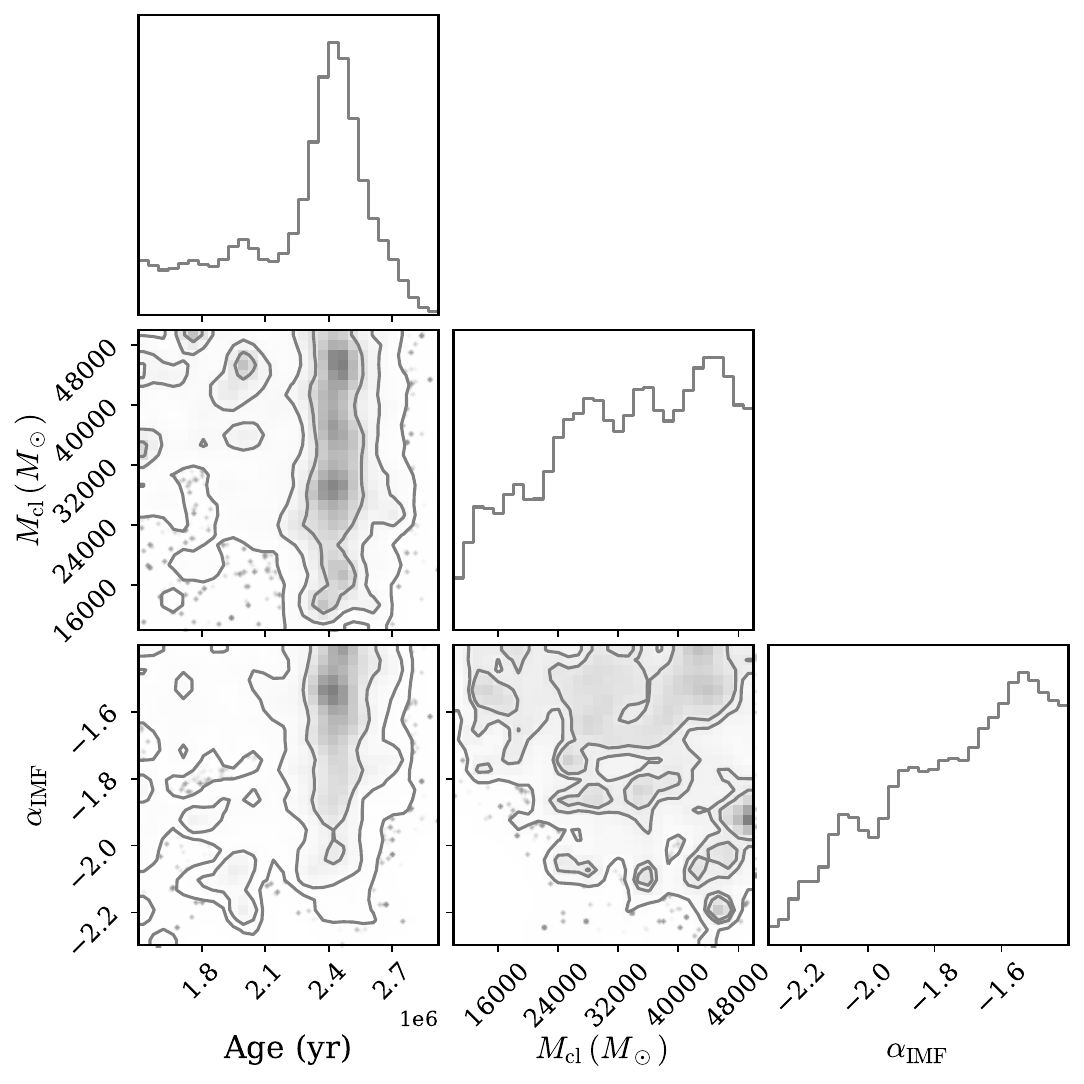}
     \caption{GENEC posterior distributions at $Z=0.02$ using a flux-only likelihood.}
    \label{Fig_fluxonly_Ekstrom}
\end{figure}

\begin{figure}
  \centering
  \includegraphics[width=\hsize]{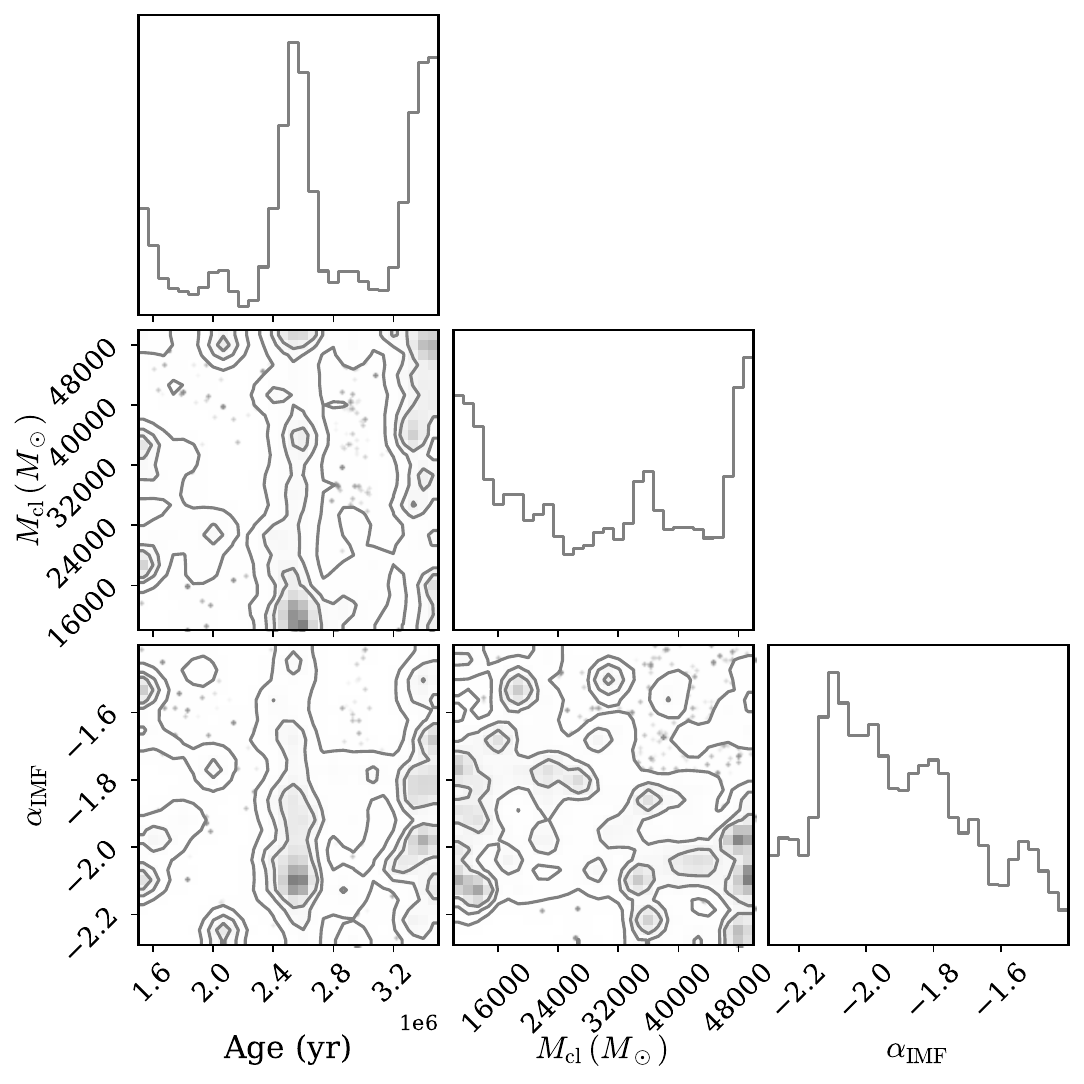}
     \caption{PARSEC posterior distributions at $Z=0.014$ using only the flux term of the likelihood.}
    \label{Fig_fluxonly_PARSEC}
\end{figure}

\begin{figure}
  \centering
  \includegraphics[width=\hsize]{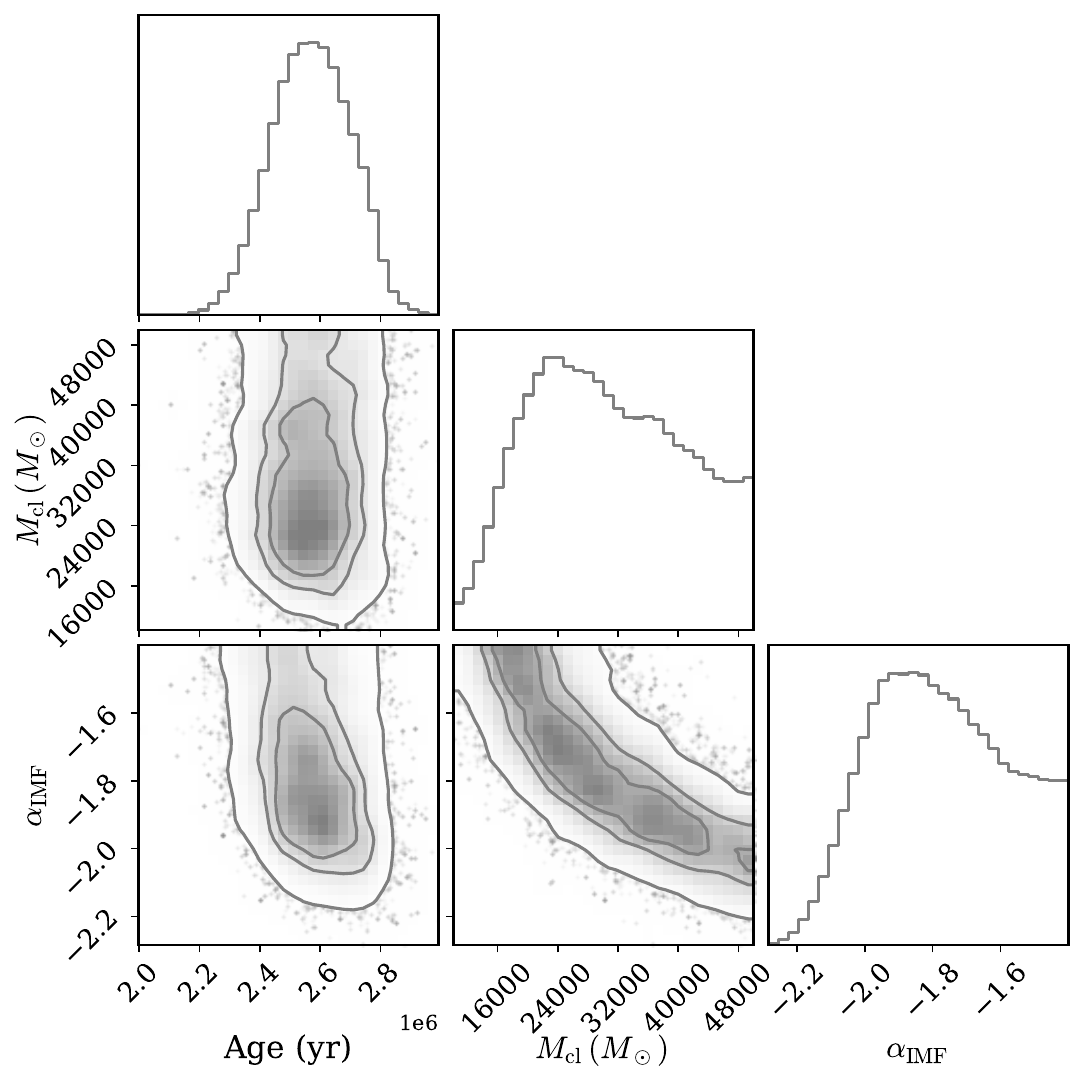}
     \caption{PARSEC posterior distributions at $Z=0.014$ for a counts-only likelihood. Note that ages are more restricted by synthetic radio-counts in the PARSEC models, presumably because of the early appearance of WC stars in the simulated clusters.}
    \label{Fig_countsonly_PARSEC}
\end{figure}

\end{appendix}

\end{document}